\documentclass[]{article}
\usepackage[margin=1in]{geometry}

\usepackage[figuresright]{rotating}

\usepackage[utf8]{inputenc}
\usepackage{graphicx}
\usepackage{xcolor, soul}
\usepackage{colortbl}
\usepackage{multirow,multicol}
\usepackage{bigints}
\usepackage{eurosym}
\usepackage{algorithm}
\usepackage[norelsize,ruled,titlenotnumbered,algo2e]{algorithm2e}
\usepackage{authblk}
\usepackage{stfloats}

\usepackage{natbib}
\usepackage[font=small,labelfont=bf]{caption}

\usepackage[british]{babel}

\usepackage{setspace}
\usepackage{sectsty}
\sectionfont{\large\uppercase}

\SetAlFnt{\small}
\SetAlgoCaptionSeparator{}
\DontPrintSemicolon
\SetAlgoLined
\SetAlgoInsideSkip{medskip}
\SetKwFor{For}{For}{}{end}
\SetKwFor{ForEach}{For each}{}{end}
\SetKwIF{If}{ElseIf}{Else}{If}{}{else if}{else}{end}

\author{}
\author[1]{D. Rios Insua}
\author[1]{A. Couce-Vieira}
\author[2]{J.A. Rubio}
\author[3]{W. Pieters}
\author[3]{K. Labunets}
\author[4]{D. G. Rasines}

\affil[1]{Instituto de Ciencias Matemáticas, Consejo Superior de Investigaciones Científicas, Madrid, Spain, david.rios@icmat.es, aitor.couce@icmat.es}
\affil[2]{Analysis, Security and Systems Group, Complutense University of Madrid, Spain, joseantonio.rubio@fdi.ucm.es}
\affil[3]{Safety \& Security Science Group, Delft University of Technology, Delft, The Netherlands, w.pieters@tudelft.nl, k.labunets@tudelft.nl}
\affil[4]{Department of Mathematics, Imperial College, London, UK, daniel.garcia-rasines16@imperial.ac.uk}

\title{An Adversarial Risk Analysis Framework for Cybersecurity}

\date{}

\begin{document}

\maketitle

\begin{abstract}
\noindent Cyber threats affect all kinds of organisations. Risk analysis is an essential methodology for cybersecurity as it allows organisations to deal with the cyber threats potentially affecting them, prioritise the defence of their assets and decide what security controls should be implemented. Many risk analysis methods are present in cybersecurity models, compliance frameworks and international standards. However, most of them employ risk matrices, which suffer shortcomings that may lead to suboptimal resource allocations. We propose a comprehensive framework for cybersecurity risk analysis, covering the presence of both adversarial and non-intentional threats and the use of insurance as part of the security portfolio. A case study illustrating the proposed framework is presented, serving as template for more complex cases.

\end{abstract}

%%%%%%%%%%%%%%%%%%%%%%%%%%%%%%%%%%%%%%%%%%%%%
\section{Introduction}
\label{sec:intro}

\noindent At present, all kinds of organisations are being critically impacted by cyber threats \cite{Anderson2008,Cyberwarfare2013}. The Cyberspace is even described as a fifth military operational space in which movements by numerous countries are common \cite{LeakSource2014}. Risk analysis is a fundamental methodology to help manage such issues \cite{Cooke&Bedford2001}. With it, organizations can assess the risks affecting their assets and what security controls should be implemented to reduce the likelihood of such threats and/or their impacts, in case they are produced.

Numerous frameworks have been developed to screen cybersecurity risks and support resource allocation, including CRAMM \cite{CRAMM2003}, ISO 27005 \cite{ISO27005}, MAGERIT \cite{Magerit2012}, EBIOS \cite{ANSSI1995}, SP 800-30 \cite{NIST2012}, or CORAS \cite{CORAS2001}. Similarly, several compliance and control assessment frameworks, like ISO 27001 \cite{ISO27001}, Common Criteria \cite{CC2012}, or CCM \cite{CSA2016} provide guidance on the implementation of cybersecurity best practices. These standards suggest detailed security controls to protect an organisation's assets against risks. They have virtues, particularly their extensive catalogues of threats, assets and security controls providing detailed guidelines  for the implementation of countermeasures and the protection of digital assets. Even though, much remains to be done regarding cybersecurity risk analysis from a methodological point of view. Indeed, a detailed study of the main approaches to cybersecurity risk management and control assessment reveals that they often rely on risk matrices, with shortcomings well documented in Cox \cite{Cox2008}: compared to more stringent methods, the qualitative ratings in risk matrices (likelihood, severity and risk) are more prone to ambiguity and subjective interpretation, and very importantly for our application area, they systematically assign the same rating to quantitatively very different risks, potentially inducing suboptimal security resource allocations. Hubbard and Seiersen \cite{hubbard} and Allodi an Massacci \cite{allodi} provide additional views on the use of risk matrices in cybersecurity. Moreover, with counted exceptions like IS1 \cite{HMG}, those methodologies do not explicitly take into account the intentionality of certain threats. Thus, ICT owners may obtain unsatisfactory results in relation with the proper prioritisation of risks and the measures they should implement.

In this context, a complementary way for dealing with cyber risks through risk transfer is emerging: cyber insurance products, of very different nature and not in every country, have been introduced in recent years by companies like AXA, Generali, Allianz, or Zurich. However, cyber insurance has yet to take off \cite{Survey2017}.

In this paper we propose a more rigorous framework for risk analysis in cybersecurity. We emphasise adversarial aspects for better prediction of threats as well as include cyber insurance.   Sect.~\ref{sec:araframe} presents our framework, supported by a case study in Sect.~\ref{sec:casestudy}. We conclude with a brief discussion.

%%%%%%%%%%%%%%%%%%%%%%%%%%%%%%%%%%%%%%%%%%%%%
\section{A cybersecurity adversarial risk analysis framework}
\label{sec:araframe}

\noindent We introduce our integrated risk analysis approach to facilitate resource allocation decision-making regarding cybersecurity. Our aim is to improve current cyber risk analysis frameworks, introducing dynamic schemes that incorporate all relevant parameters, including decision-makers' preferences and risk attitudes \cite{Clemen2013} and the intentionality of adversaries. Moreover, we introduce decisions concerning cyber insurance adoption to complement other risk management decisions through risk transfer. Fielder et al. \cite{fielder2016} review and introduce various approaches to cyber security investment, which cover optimisation and/or game theoretic elements, under strong common knowledge assumptions. Our framework combines optimisation with an adversarial risk analysis (ARA) approach to deal with adversarial agents.

We present the framework stepwise, analysing the elements involved progressively. We describe the models through influence diagrams (ID) and bi-agent influence diagrams (BAID) \cite{Banks2015} detailing the relevant elements:  assets, threats, security controls, costs and benefits. We provide a brief verbal description of the diagrams introduced and a generic mathematical formulation at each step.

%%%%%%%%%%%%%%%%%%%%%%%%%%%%%%%%%%%%%%%%%%%%%
\subsection{System performance evaluation}
\label{subsec:sysperformance}

\begin{figure}
	\centering
	\includegraphics[scale=.8]{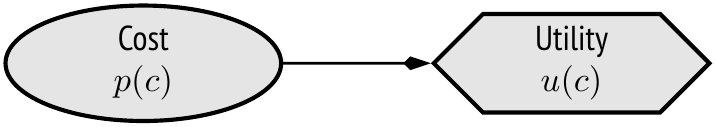}
	\caption{Basic influence diagram for performance evaluation.}
	\label{basicspe}
\end{figure}

\noindent Fig.~\ref{basicspe} describes the starting outline for a system under study. Costs associated with system operation over the relevant planning period are indicated by $c$. Such costs are typically uncertain, modelled with a probability distribution $p(c)$. We introduce a utility function $u(c)$ \cite{Ortega2017} over costs to cater for risk attitudes.
The evaluation of system performance under normal conditions, i.e. in absence of relevant cyber incidents, is based on its associated expected utility \cite{French&Rios2000}
\[\psi_n = \int u(c) \, p(c) \ dc .\]
\begin{figure}
	\centering
	\includegraphics[scale=.8]{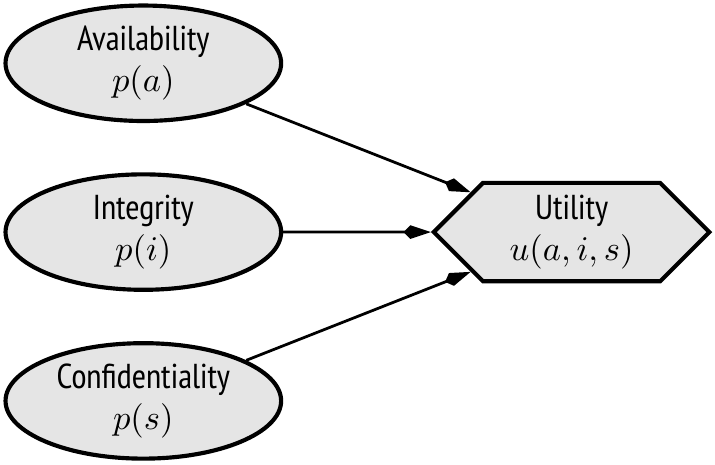}
	\caption{Cybersecurity attributes for performance evaluation.}
	\label{ciaspe}
\end{figure}
This basic scheme can be sophisticated in several directions. For example, there could be several performance functions. A typical case is to consider attributes concerning information availability ($a$), integrity ($i$) and confidentiality ($s$) \cite{Mowbray2013}, Fig.~\ref{ciaspe}. These nodes could be, in turn, antecessors of the cost node. We use $p(a, i, s)$ as the distribution modelling uncertainty about system performance. If $u(a, i, s)$ represents the  corresponding multi-attribute utility, the expected utility would be
\[\psi_{n} = \iiint u(a, i, s) \, p(a, i, s) \ da \, di \, ds .\]
\noindent We use $p(a, i, s)$ if interrelationships between such attributes are expected. If this were not so, e.g.\ in the case of independence, we would describe graphically the model as in Fig.~\ref{ciaspe}, through
 \[ p(a, i, s) = p(a) \, p(i) \, p(s) .\]

%%%%%%%%%%%%%%%%%%%%%%%%%%%%%%%%%%%%%%%%%%%%%%%
\subsection{Cybersecurity risk assessment}
\label{subsec:ranacyber}

\begin{figure}
	\begin{center}
		\includegraphics[scale=.8]{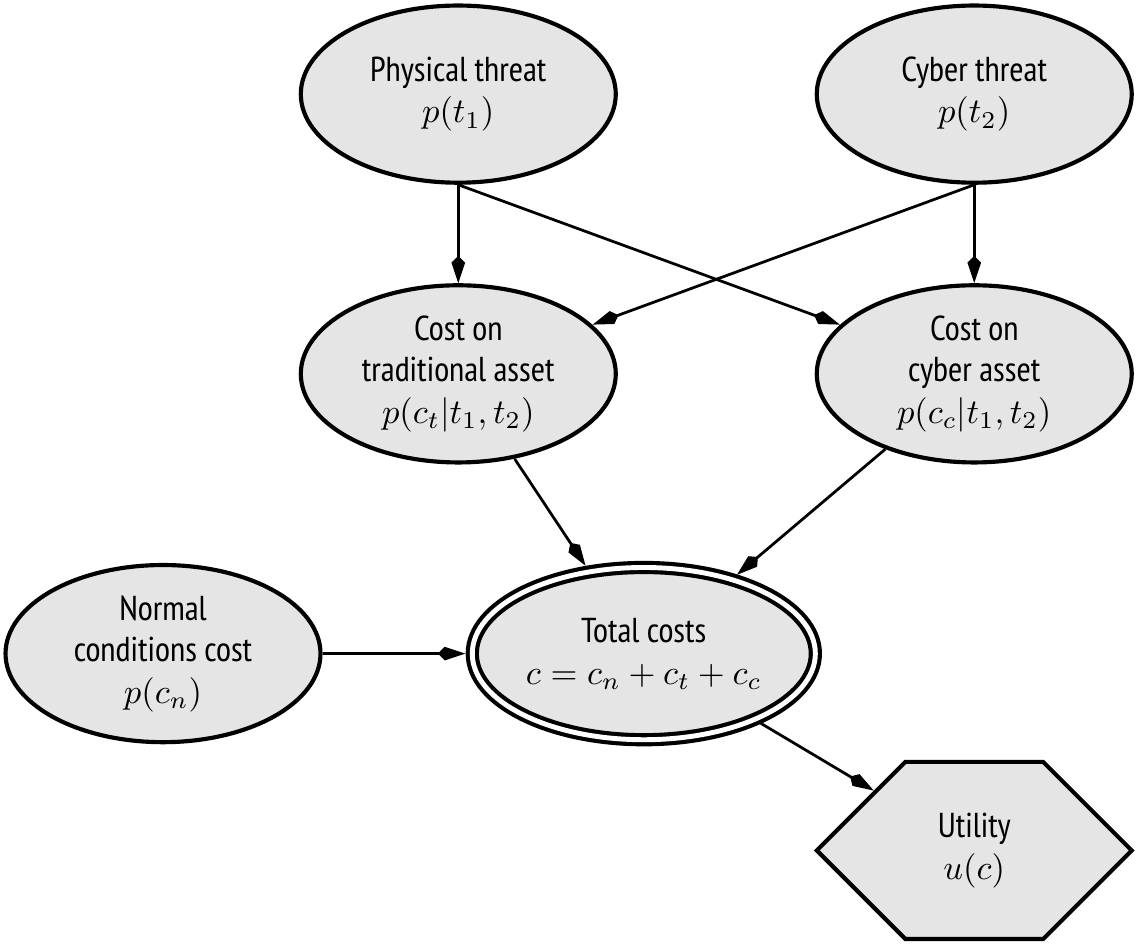}
		\caption{Risk assessment in cybersecurity.}
		\label{racyber} 
	\end{center}
\end{figure}

\noindent Adopting the basic scheme in Fig.~\ref{basicspe}, on which we focus to simplify the exposition, we consider the problem of cybersecurity risk assessment in Fig.~\ref{racyber}. For instance, consider a model with just two threats, one of them ($t_1$) physical (e.g., fire) and another one ($t_2$) cyber (e.g., DDoS attack\footnote{A distributed denial of service (DDoS) is a network attack consisting of a high number of infected computers flooding with network traffic a victim computer or network device, making it inaccessible.}). Both $t_1$ and $t_2$ are random variables. We also consider two types of assets, one traditional (e.g., facilities) and the other cyber (e.g., computers). Impacts over these assets are, respectively, $c_t$ and $c_c$ and, typically, uncertain. If there is a relationship between them given either threat, the corresponding model would be of the form
\[p(c_t, c_c | t_1, t_2) \, p(t_1, t_2), \]
where $ p(t_1, t_2) $ describes the probability of the threats happening, and $ p(c_t, c_c | t_1, t_2) $ describes the probability over asset impacts, given the eventual occurrence of threats. Costs are added at the total cost node $c$, which aggregates those under normal circumstances  with those due to the incidents. Then, the expected utility taking into account the threats and specific dependencies in Fig.~\ref{racyber} would be
\[
\psi_r = \int \dots \int \
u(c_n + c_t + c_c) \, p(c_n) \, p(c_{t} | t_{1}, t_{2}) \, p(c_{c} | t_{1}, t_{2}) \, p(t_{1}) \, p(t_{2}) \
dt_2 \, dt_1 \, dc_c \, dc_t \, dc_n.
\]
\noindent We have assumed that consequences are additive, but we could have a generic utility $ u(c_n, c_c, c_t) $. Finally, we evaluate the loss $ \psi_n - \psi_r $ in expected utility considering the threats against that under normal conditions. When it is sufficiently large, incidents are expected to harm the system significantly and we should manage such risks.

The model can be extended to include a bigger number of threats and assets, as well as additional types of costs. Finally, several utility nodes could be incorporated to describe the preferences of multiple stakeholders.

%%%%%%%%%%%%%%%%%%%%%%%%%%%%%%%%%%%%%%%%%%%%%
\subsection{Risk mitigation in cybersecurity risk management}
\label{subsec:rmancyber}

\begin{figure}
	\begin{center}
		\includegraphics[scale=0.8]{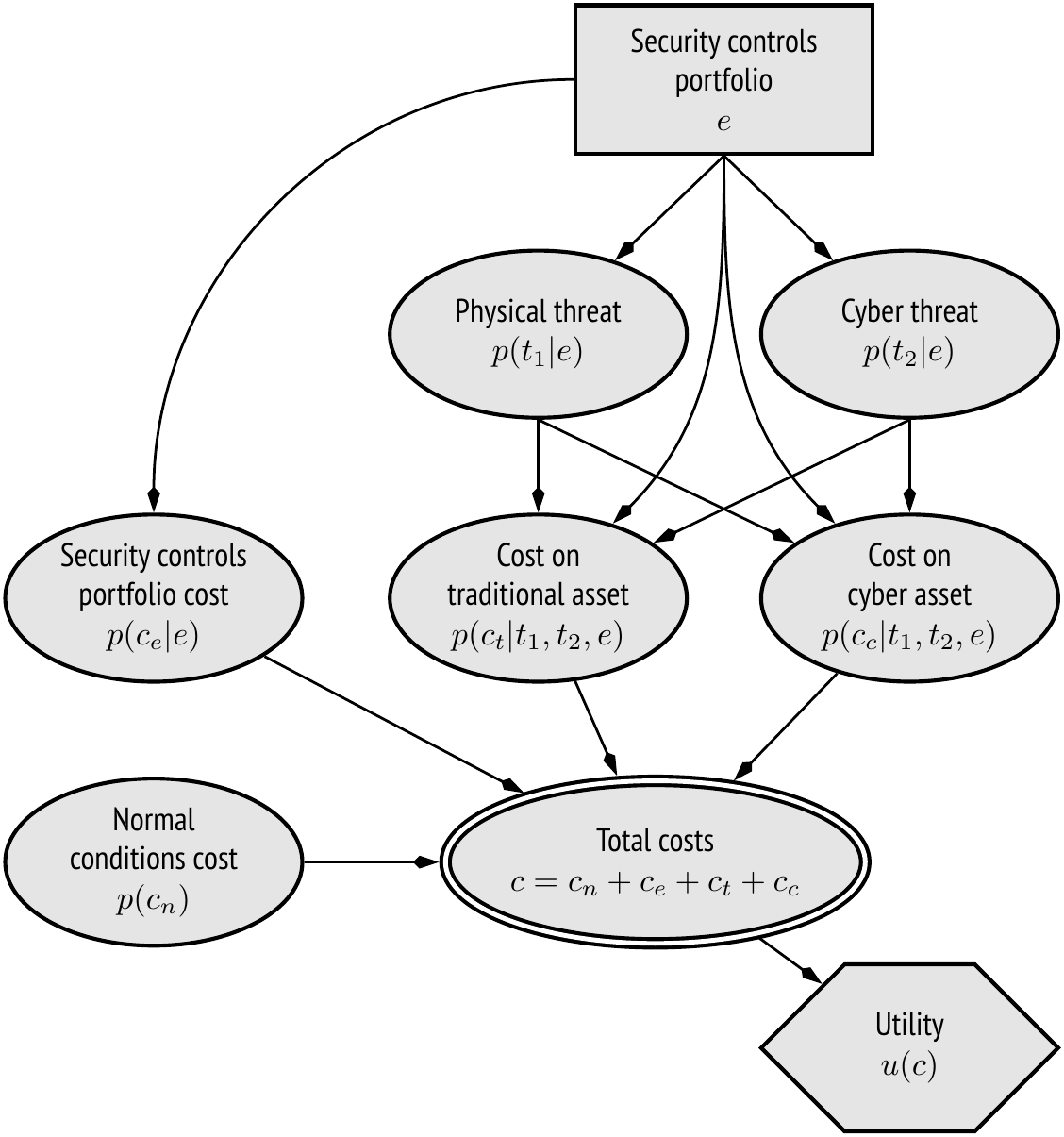}
		\caption{Risk assessment of cybersecurity controls.}
		\label{rmcyber} 
	\end{center}
\end{figure}

\noindent The next step adds security controls to the model. We introduce a portfolio of them to reduce the likelihood of threats and/or their impact (Fig.~\ref{rmcyber}). Examples of controls include firewalls, employee training, or making regular backups.
For simplicity, in Fig.~\ref{rmcyber} we assume that all controls influence over all events and impacts. It will not always be so: a fire detector makes less harmful, but not less likely, a fire; resource accounting mechanisms \cite{DDoS} managing access based on user privileges, make less likely a DDoS, but usually not less harmful. Node $e$ describes the portfolio of controls, whose cost we model through the distribution $p(c_e | e)$. Controls might influence on threat likelihoods $p(t_1 | e)$ and $p(t_2 | e)$, as well as on asset impact likelihoods $p(c_t | t_1, t_2, e)$ and $p(c_c |t_1, t_2, e)$. All costs are aggregated in the total cost node $c$, under appropriate additivity assumptions.

In this case, the expected utility when portfolio $e$ is implemented is
\[
\psi (e) = \int ...\int \
u(c_n + c_e + c_t + c_c) \, p(c_n) \, p(c_e | e) \, p(c_t | t_1, t_2, e) \, p(c_c | t_1, t_2, e) \, p(t_1 | e) \, p(t_2 | e) \ dt_2 \, dt_1 \, dc_e \, dc_t \, dc_c \, dc_n .
\]
\noindent We would then look for the maximum expected utility portfolio solving for
\[\psi_{e}^{*} = \max_{e \in E} \psi (e),\]
\noindent being $E$ the set of feasible portfolios. Based on the available controls, we define portfolios that meet different constraints which may be economic (e.g., not exceeding a budget), legal (e.g., complying with data protection laws), logistic or physical.

%%%%%%%%%%%%%%%%%%%%%%%%%%%%%%%%%%%%%%%%%%%%%
\subsection{Cyber insurance in cybersecurity risk management}
\label{subsec:cyberins}

\begin{figure}
	\begin{center}
		\includegraphics[scale=0.8]{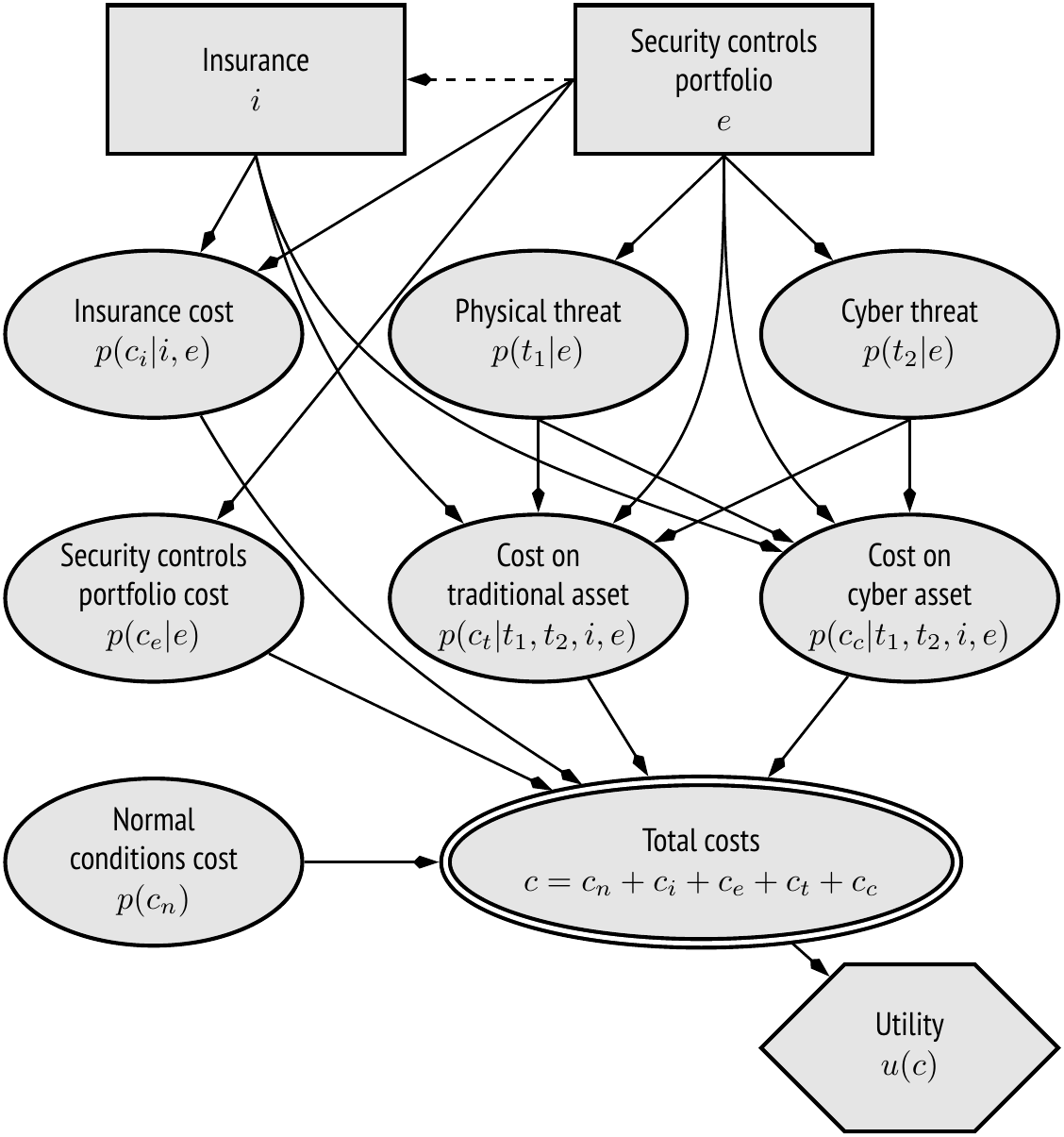}
		\caption{Cyber insurance in cybersecurity risk assessment.}
		\label{rtcyber}
	\end{center}
\end{figure}

\noindent As a relevant element of increasing interest, we introduce cyber insurance. Its costs will typically depend on the implemented portfolio of controls, as in Fig.~\ref{rtcyber}: the better such portfolio is, the lower the premium will be. This cost will also depend on the assets to be protected. We could include the insurance within the portfolio of controls; however, it is convenient to represent them separately, since premiums will typically depend on the controls deployed.
Decision node $i$ describes the cyber insurance adopted, with entailed costs $c_i$ with probability $p(c_i | i, e)$, although they will usually be deterministic. In addition, insurance and security controls will typically affect impacts, modelled through $p(c_t | t_1, t_2, e, i)$ and $p(c_c | t_1, t_2, e, i)$. Costs are aggregated in the total cost node $c$. The expected utility when portfolio $e$ is implemented together with insurance $i$ is
\begin{multline*}
\psi (e, i) = \int \dots \int \
u(c_n + c_i + c_e + c_t + c_c) \, p(c_n) \, p(c_i | i, e) \, p(c_e | e) \, p(c_t | t_1, t_2, e, i) \, p(c_c | t_1, t_2, e, i) \, \times \\ \times \, p(t_1 | e) \, p(t_2 | e) \ dt_2 \, dt_1 \, dc_c \, dc_t \, dc_e \, dc_i \, dc_n.
\end{multline*}
\noindent We seek the maximum expected utility portfolio-insurance pair through
\[\max_{e \in E , i\in I} \psi (e, i),\]
where $I$ represents the insurance catalogue. The pair $(e,i)$ could be further restricted jointly, e.g., by a common budget constraint or legal requirements.

%%%%%%%%%%%%%%%%%%%%%%%%%%%%%%%%%%%%%%%%%%%%%
\subsection{Adversarial risk analysis in cybersecurity}
\label{subsec:aracyber}

\noindent As discussed previously, intentionality is a key factor when analysing certain cyber threats. We shall use ARA \cite{Banks2015} to model the intentions and strategic behaviour of adversarial cyber threats. Under ARA, the attacker has his own utility function $u_A$, seeking to maximise the effectiveness of his attack. This paradigm is applicable to multiple types of strategic interactions between attackers and defenders. Two of them are specially relevant in cybersecurity. First, the sequence \emph{defence-attack}, in which the Defender deploys her security controls and the Attacker is able to observe them prior to attacking. Second, the sequence \emph{defence-attack-defence}, in which the Defender deploys her preventive controls, then the Attacker observes them to decide his attack and, finally, the Defender recovers from the attack, should it be successful.

%%%%%%%%%%%%%%%%%%%%%%%%%%%%%%%%%%%%%%%%%%%%%
\subsubsection{Defence-attack model}
\begin{figure}
	\begin{center}
		\includegraphics[scale=0.8]{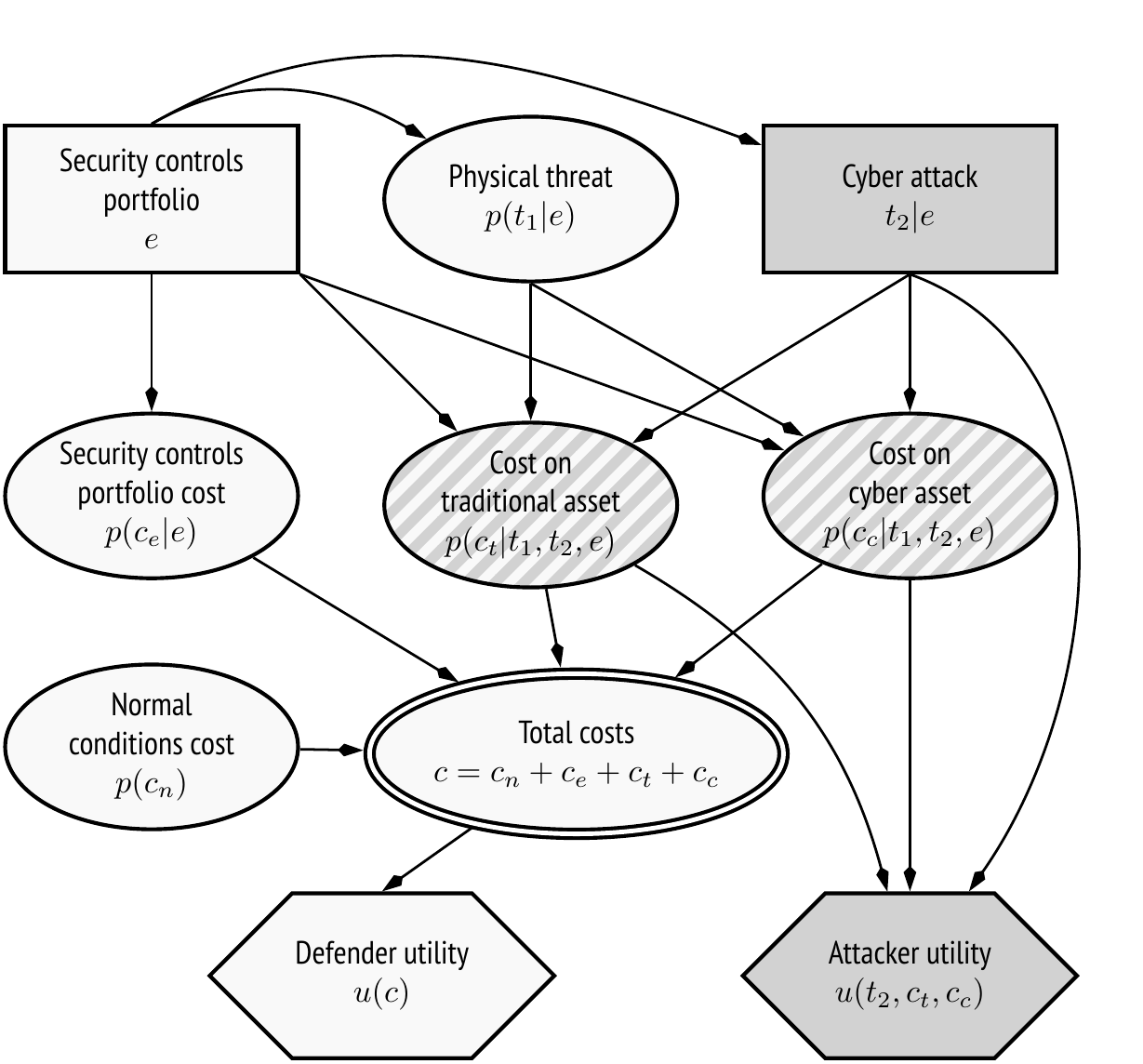}
		\caption{Adversarial risk analysis in cybersecurity: defence-attack problem.}
		\label{figarada1}
	\end{center}
\end{figure}

\noindent The original examples, Figs.~\ref{racyber} and~\ref{rmcyber} evolve into Fig.~\ref{figarada1}, modelling an adversarial case through a BAID with a Defender and an Attacker: physical threat $t_{1}$ remains unintentional whereas cyber threat $t_{2}$ becomes adversarial through a decision node for the Attacker, who needs to decide whether or not to launch an attack to his benefit. It corresponds to a sequential defence-attack model \cite{Banks2015}.
\begin{figure}
	\begin{center}
		\includegraphics[scale=0.8]{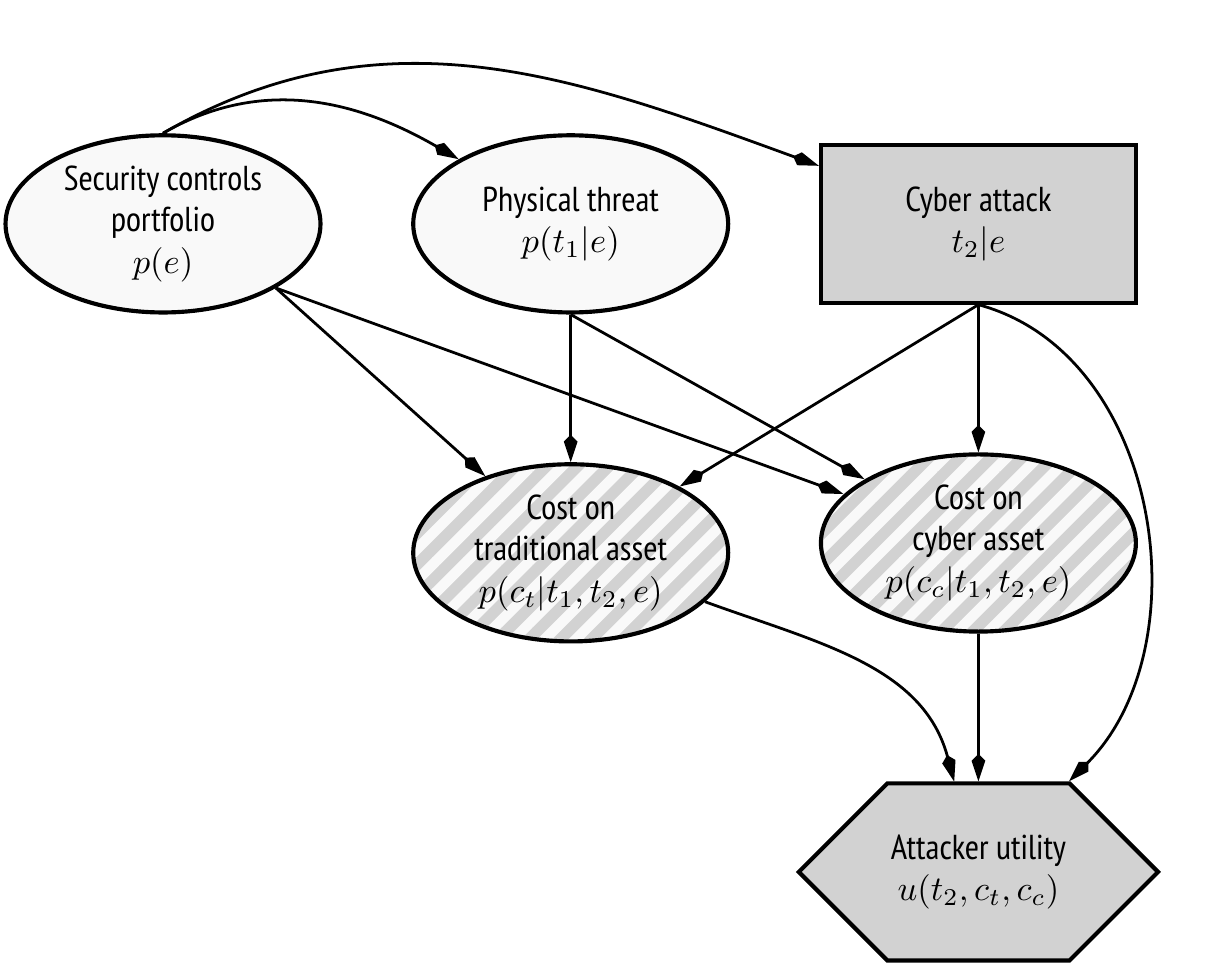}
		\caption{Attacker problem in the defence-attack model.}
		\label{figarada2}
	\end{center}
\end{figure}
The Defender problem is described in Fig.~\ref{rmcyber}. Its resolution was covered in Sect.~\ref{subsec:rmancyber}. There, the cyber attack is described probabilistically through $p(t_{2} | e)$, which represents the probability that the Defender assigns to cyber threat $t_{2}$ materialising, had portfolio $e$ been adopted. However, the strategic nature of this problem, Fig.~\ref{figarada1}, requires the analysis of the Attacker decision about which attack to perform. Under the ARA defence-attack paradigm, the Defender should analyse the Attacker strategic problem in Fig.~\ref{figarada2}.

Specifically, given portfolio $e$, and assuming that the Attacker maximizes expected utility, the Defender would compute for each attack $t_{2}$, the expected utility for the Attacker
\[
\psi_A (t_2 | e) =
\iiint u_A (t_2, c_t, c_c) \, p_A (c_t | t_1, t_2, e) \, p_A (c_c |t_1, t_2, e) \, p_A (t_1 | e) \ dt_1 \, dc_c \, dc_t,
\]
where $u_A$ and $p_A$ are, respectively, the utilities and probabilities of the Attacker. The Defender must then find the attack maximising the Attacker's expected utility,
\[\max_{t_2\in T_2} \psi_A (t_2 | e) ,\]
where $T_2$ is the attack set.

However, the Defender will not typically know $u_{A}$ and $p_{A}$. Suppose we are capable of modelling her uncertainty about them with random probabilities $P_{A}$ and a random utility function $U_{A}$ \cite{Banks2015}. Then, the optimal random attack given $e$ is
\[
T^{*}_{2} (e) =
\arg\max_{t_2 \in T_2}
\iiint U_A (t_2, c_t, c_c) \, P_A (c_t | t_1, t_2, e) \, P_A (c_c | t_1, t_2, e) \, P_A (t_1 | e) \ dt_1 \, dc_c \, dc_t.
\]
\noindent Finally, the distribution over attacks that we were looking for satisfies
\[p (t_2 | e) = P \big(T^*_2 (e) = t_2 \big) ,\]
\noindent assuming that $T_2$ is discrete (e.g., when referring to attack options), and, similarly, if they are continuous (e.g., when referring to attack efforts). Such distribution could be estimated through Monte Carlo (MC) simulation as in Algorithm~\ref{algo1} (Appendix), where the distribution of random utilities and probabilities is designated by
\[
F =
\Big(
U_{A} (t_2, c_t, c_c), P_A (c_t | t_1, t_2, e), P_A (c_c | t_1, t_2, e), P_A (t_1 | e)
\Big)
\]

%%%%%%%%%%%%%%%%%%%%%%%%%%%%%%%%%%%%%%%%%%%%%
\subsubsection{Defence-attack-defence model}

\noindent As mentioned, cybersecurity risk management also comprises reactive measures that can be put in place to counter an attack, should it happen. Therefore, we split the security portfolio into two groups: preventive security controls $e_p$ and reactive security controls $e_r | t_2$. This corresponds to a sequential defence-attack-defence model \cite{Banks2015} in which the first move is by the Defender (preventive portfolio $e_p$), the second is by the Attacker (attack after observing preventive controls, $t_2|e_p$) and the third one is by the Defender (reactive portfolio $e_r|t_2$).
\begin{figure}
	\begin{center}
		\includegraphics[scale=0.8]{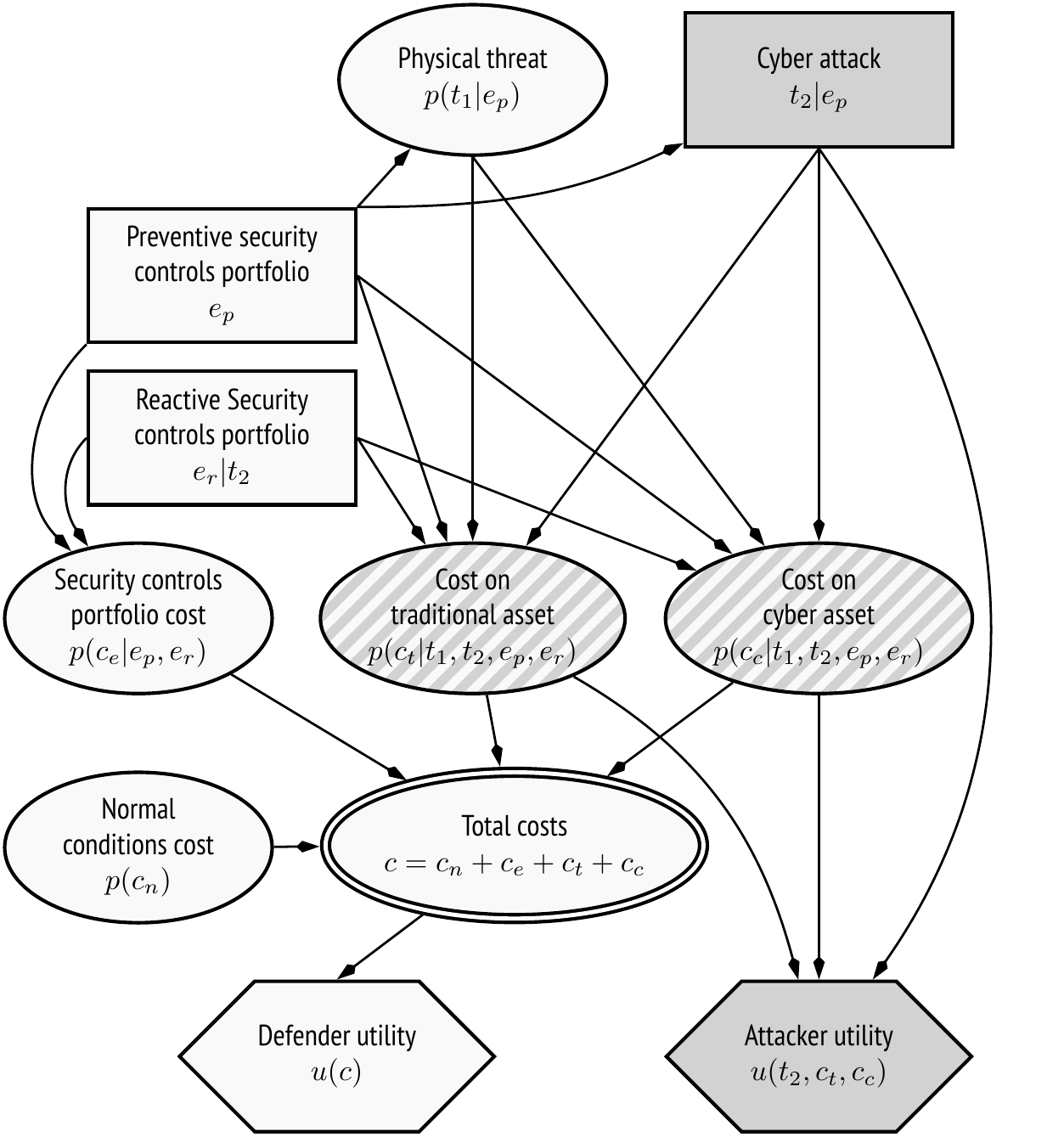}
		\caption{Adversarial risk analysis in cybersecurity: defence-attack-defence problem..}
		\label{figaradad1}
	\end{center}
\end{figure}
The Defender problem is solved similarly to Sect.~\ref{subsec:rmancyber}, reflecting changes caused by splitting
the security control node. Specifically, the expected utility when portfolio $ e=(e_p, e_r) $ is implemented is
\begin{multline*}
\psi (e) = \int ...\int
u(c_n + c_e + c_t + c_c) \, p(c_n) \, p(c_e | e_p, e_r) \, p(c_t | t_1, t_2, e_p, e_r) \, p(c_c | t_1, t_2, e_p, e_r) \, \times \\ \times \, p(t_1 | e_p) \, p(t_2 | e_p) \ dt_2 \, dt_1 \, dc_c \, dc_t \, dc_e \, dc_n .
\end{multline*}
\noindent We would then look for the maximum expected utility portfolio
\[(e_p^*, e_r^*) = \underset{(e_p, e_r) \in E_p \times E_r}{\arg\max} \psi(e_p, e_r),\]
\noindent where $E_p$ and $E_r$, respectively, define constraints for preventive and reactive portfolios, some of which could be joint. The Attacker problem providing $p(t|e_p)$ would be solved in a similar fashion than in the defence-attack case.

%%%%%%%%%%%%%%%%%%%%%%%%%%%%%%%%%%%%%%%%%%%%%
\section{A case study template}
\label{sec:casestudy}
We illustrate our framework for cybersecurity risk analysis with a defend-attack case study, which can serve as a template for more complex cases. The Defender is an SME dedicated to document management with 60 people and 90 computers. A cyber attack might affect, mainly, the online document management service. For confidentiality reasons, the number of relevant issues has been simplified and data conveniently masked. This simplification will allow us to better illustrate key modelling concepts and the overall scheme to follow for other case studies. Moreover, we include uncertain phenomena in which data are available and others in which it is not and, thus, we shall need to rely on expert judgement \cite{dias2018}. Prices and rates refer to Spain, where the incumbent SME is located.

In essence, we first structure the problem identifying assets, threats and security controls. The later may have implementation costs in exchange of reducing the threat likelihoods and/or eventual impacts. Subsequently, we assess the impacts that may have an effect on asset values to find the optimal risk management portfolio. Since adversarial threats are included, we also model the Attacker decision problem.  Indeed, in this case there is a single potential Attacker which contemplates a DDoS attack with the objective of disrupting the SME services, causing an operational disruption and reputational damage with the consequent loss of customers, which would head to competitors, besides incurring in contractual penalties potentially affecting its continuity. Then, we simulate from this problem to obtain the attack probabilities, which feed back the Defender problem. In this way an optimal defence can be obtained. We consider a one-year planning horizon.

The problem we focus is on finding the optimal security portfolio and insurance product for the company, in the sense of maximising expected utility.
Other formulations are discussed in Sect.~\ref{subsec:further}.

%%%%%%%%%%%%%%%%%%%%%%%%%%%%%%%
\subsection{Problem structuring}
\label{subsec:problemstructuring}
\begin{figure*}[h]
	\begin{center}
		\includegraphics[scale=0.7]{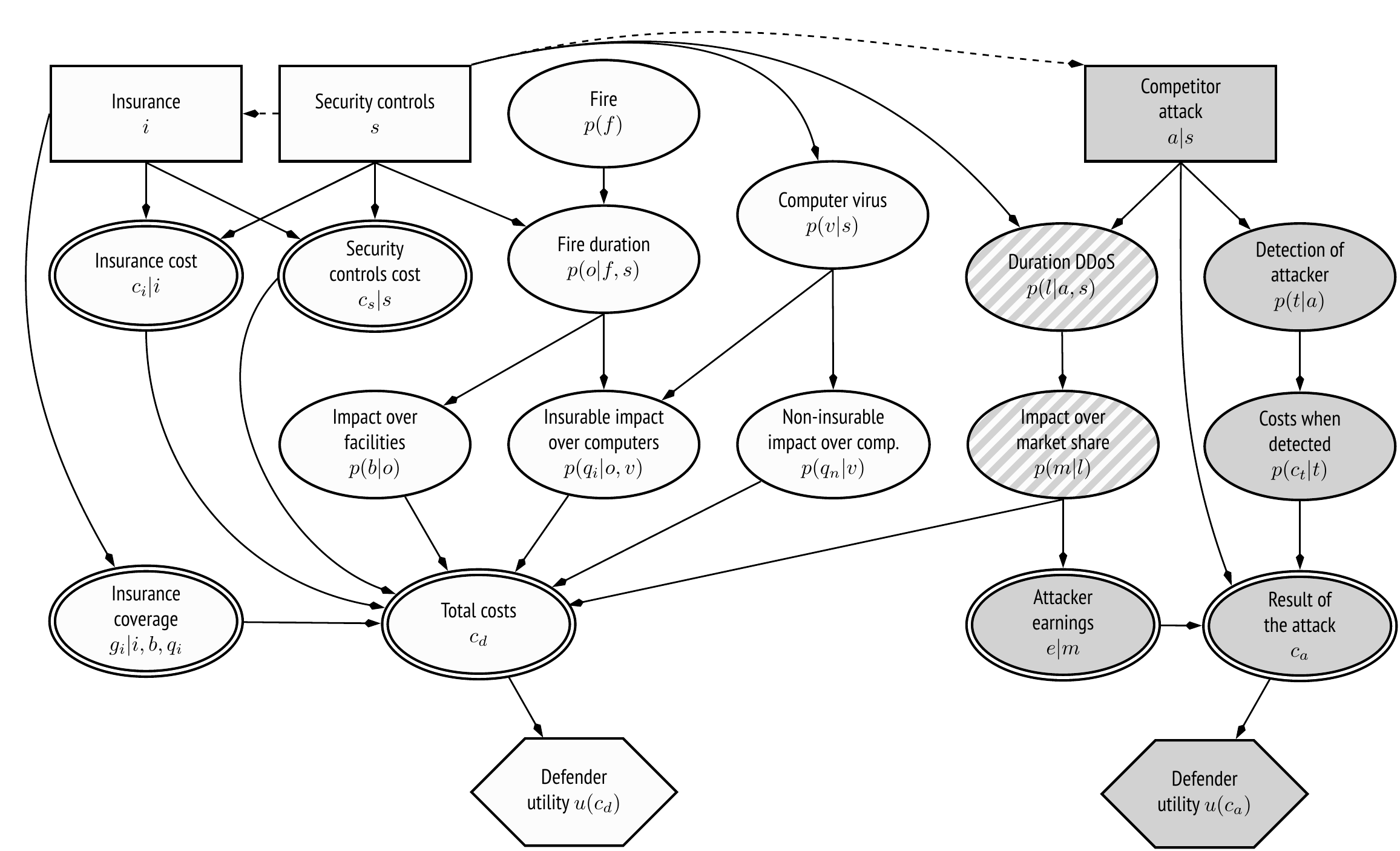}
		\caption{Case problem structure as a BAID.}
		\label{fig8}
	\end{center}
\end{figure*}

We structure the problem through the BAID in Fig.~\ref{fig8}. Lighter nodes refer to issues concerning solely the Defender; darker nodes refer to issues relevant only for the Attacker; nodes with an stripped background affect both. Should there be several attackers, we would use more background patterns or colours.
Arcs have the same interpretation as in \cite{shachter1986evaluating}. The only non-standard arc is that linking the security controls node and the attack node, meaning that the Attacker will implement his action once he identifies the controls adopted by the Defender.

%%%%%%%%%%%%%%%%%%%%%%%%%%%%%%%%
\subsubsection{Assets}	
 We first identify the Defender assets at risk. We could obtain them from catalogues like those of the methodologies mentioned in the Introduction. Here we consider:
\begin{itemize}
	\item \emph{Facilities}: Offices potentially affected by threats.  Without them, the organisation could not operate.
	\item \emph{Computer equipment}: The data centre and workstations essential for this organisation. Should they be affected, costs could be substantial.
	\item \emph{Market share}, directly impacting the company profits.
\end{itemize}

Other assets not considered in this case include the company's development software, business information, the mobile computing elements or the staff.

%%%%%%%%%%%%%%%%%%%%%%%%%%%%%%%%%%%%%%%%%%%%%%
\subsubsection{Non-intentional threats}
 We consider threats over the identified assets deemed relevant and having non-intentional character. This may include threats traditionally insurable as well as new ones potentially cyber insurable. We use a simplification of the catalogues in the methodologies in the introduction:
\begin{itemize}
	\item \emph{Fire}: It may affect facilities, as well as computers, which could even destroyed. No impact over market share is contemplated, as the organisation has a backup system. We assume that a fire can occur only by accident, not considering the possibility of sabotage.
	\item \emph{Computer virus}: Aimed at disrupting normal operations of computer systems. We consider this threat non-intentional, as most viruses propagate ubiquitously: their occurrence tends to be random from the defender perspective. It may degrade computer performance.
\end{itemize}
We model each threat with a probabilistic node associated with the Defender problem. Other non-intentional threats, not considered here, could be water damage, power outages
 or employee errors.

%%%%%%%%%%%%%%%%%%%%%%%%%%%%%%%%%%%%%%%%%%%%%%%%
\subsubsection{Intentional threats}
This category may include both cyber and physical threats. Again, we may use catalogues from, e.g., MAGERIT. We should identify the corresponding attackers, as well as their attack options available. In our case, we just consider a relevant attacker.
\begin{itemize}
	\item \emph{Competitor attack}: Our competitor may attempt a DDoS, to undermine the availability of the Defender's site, compromising her customer services. Should it happen, it would impact negatively the Defender's market share, damaging its reputation and, consequently, loosing customers to be gained by the Attacker. The decision is whether to launch the attack and the number of attempts.
\end{itemize}

\noindent We integrate attack options into a single decision node associated with the Attacker problem. Other intentional attacks, not modelled here, could include an abuse of access privileges, launching an advanced persistent threat, insiders or bombs.

%%%%%%%%%%%%%%%%%%%%%%%%%%%%%%%%%%%%%%%%%%%%%%%%%%%%%
\subsubsection{Uncertainties affecting threats}
 We consider now those uncertainties affecting the Defender's assets.
\begin{itemize}
	\item \emph{Duration of DDoS}, will depend on the number of attacks and security controls deployed.
	\item \emph{Fire duration}, which can be reduced with an anti-fire system.
\end{itemize}

\noindent Each one is modelled with a probabilistic node. Other related uncertainties could come, e.g., from a more detailed modelling of the virus (e.g., infection probability given the OS) or the fire propagation to adjacent buildings.

%%%%%%%%%%%%%%%%%%%%%%%%%%%%%%%%%%%%%%%
\subsubsection{Attacker uncertainties}
 Additionally, we consider uncertainties that the Attacker might find relevant in his problem and affect only him.
\begin{itemize}
	\item \emph{Detection of Attacker}. If detected, his reputation would suffer and might face legal prosecution.
\end{itemize}

\noindent Each of them is modelled with a probabilistic node. Other attacker uncertainties include the number of customers affected by the DDoS or the performance of the attack platform.

%%%%%%%%%%%%%%%%%%%%%%%%%%%%%%%%%%%%%%%%%
\subsubsection{Relevant security controls}
We identify security controls relevant to counter the threats. We may use listings from the above mentioned methodologies. In our case we consider:
\begin{itemize}
	\item \emph{Anti-fire system}. It can detect a fire facilitating early mitigation.
	\item \emph{Firewall}. It protects a network from malicious traffic.
	\item Implementation of \emph{risk mitigation procedures} for cybersecurity and fire protection.
	\item \emph{Cloud-based DDoS protection}, diverting DDoS traffic from the target to a cloud-based site absorbing malicious traffic.
\end{itemize}

\noindent We associate a Defender decision node with the security controls. Other measures, not included here, could be a system resource management policy, a cryptographic data protocol or a wiring protection.

%%%%%%%%%%%%%%%%%%%%%%%%%%%%
\subsubsection{Insurance}
We also consider the possibility of purchasing insurance to transfer risk. The premium will depend on the protected assets and contextual factors such as location, company type and, quite importantly, the implemented controls. Available insurance products are in Table \ref{table2}.
\begin{table}
	{\footnotesize
		\begin{center}
			\begin{tabular}{| l | l |}
				\hline
				\textbf{Product} & \textbf{Coverage} \\
				\hline
				\emph{No insurance} & None \\
				\hline
				\parbox{7em}{\emph{Traditional \\ insurance}} & \parbox{16em}{\vspace{3pt} $80$\% of hired capital in buildings and contents; firefighters; movement of furniture. \vspace{3pt}}\\
				\hline
				\parbox{7em}{\emph{Cyber \\ insurance}} & \parbox{16em}{\vspace{3pt} $80$\% of these expenses: Those related with confidential data	violation, investigation and legal costs; losses caused by threats and extorsion; removal of computer viruses; measures related to data protection law procedures; computer fraud.} \\
				\hline
				\parbox{7em}{\vspace{3pt} \emph{Comprehensive insurance}\vspace{3pt}} & All of the above. \\
				\hline
			\end{tabular}
			\caption{\label{table2} Insurance product features.}
	\end{center}}
\end{table}
 
We associate a Defender decision node with the insurance to be contracted. As its cost depends on the implemented controls, we include the corresponding decision node as a predecessor.

%%%%%%%%%%%%%%%%%%%%%%%%%%%%%%%%%%%%%%%%%
\subsubsection{Impacts for Defender}	
 Having identified threats and assets, we present their potential impacts over the Defender's interests:
\begin{itemize}
	\item \emph{Impact over facilities}: Economic losses caused by fire over them.
	\item \emph{Impact over computers}: Economic losses caused by fire or viruses over computers. We split them into insurable impacts and non-insurable ones. We need this split to calculate the eventual insurance coverage.
	\item \emph{Impact over market share}: Costs due to market share lost.
\end{itemize}

\noindent We model each impact with a probabilistic node. We also consider the impacts associated with safeguards.
\begin{itemize}
	\item \emph{Cost of security controls} implemented by Defender.
	\item \emph{Cost of insurance} acquired.
	\item \emph{Insurance coverage}.
\end{itemize}

\noindent Finally, a node aggregates all Defender's consequences:
\begin{itemize}
	\item \emph{Total costs}: It summarises the above to establish the final monetary impact of the Defender problem.
\end{itemize}

\noindent The above cost nodes will be deterministic. Besides, we could also include other types such as corporate image or staff safety.

%%%%%%%%%%%%%%%%%%%%%%%%%%%%%%%%%%%%%%%
\subsubsection{Impacts for Attacker}
We consider the following impacts:
\begin{itemize}
	\item \emph{Attacker earnings} from increasing market share, transferred from that
 lost by the Defender.
	\item \emph{Costs when detected}, covering eventual sanctions by the regulator, legal costs as well as loss of customers and reputation, if detected.
	\item The final \emph{results of attack} combines all previous earnings and costs, as well as those of undertaking the attack, such as acquiring malicious tools or hiring hackers.
\end{itemize}

\noindent We model the costs when detected as a probabilistic node. The remaining nodes are deterministic.

%%%%%%%%%%%%%%%%%%%%%%%%%%%
\subsubsection{Preferences}
Value nodes describe how the corresponding agent evaluates consequences. We use the expected utility paradigm \cite{French&Rios2000}. We, therefore, include these nodes:
\begin{itemize}
	\item \emph{Utility of Defender:} Models the Defender preferences and risk attitudes over the total costs.
	\item \emph{Utility of Attacker:} It describes the Attacker preferences and risk attitudes.
\end{itemize}

\noindent We include a value node for each of the utility functions.

%%%%%%%%%%%%%%%%%%%%%%%%%%%%%%%%%%%%%%%%%%%%%
\subsubsection{Defender and Attacker problems}
\begin{figure*}[h]
	\centering
	\includegraphics[scale=.7]{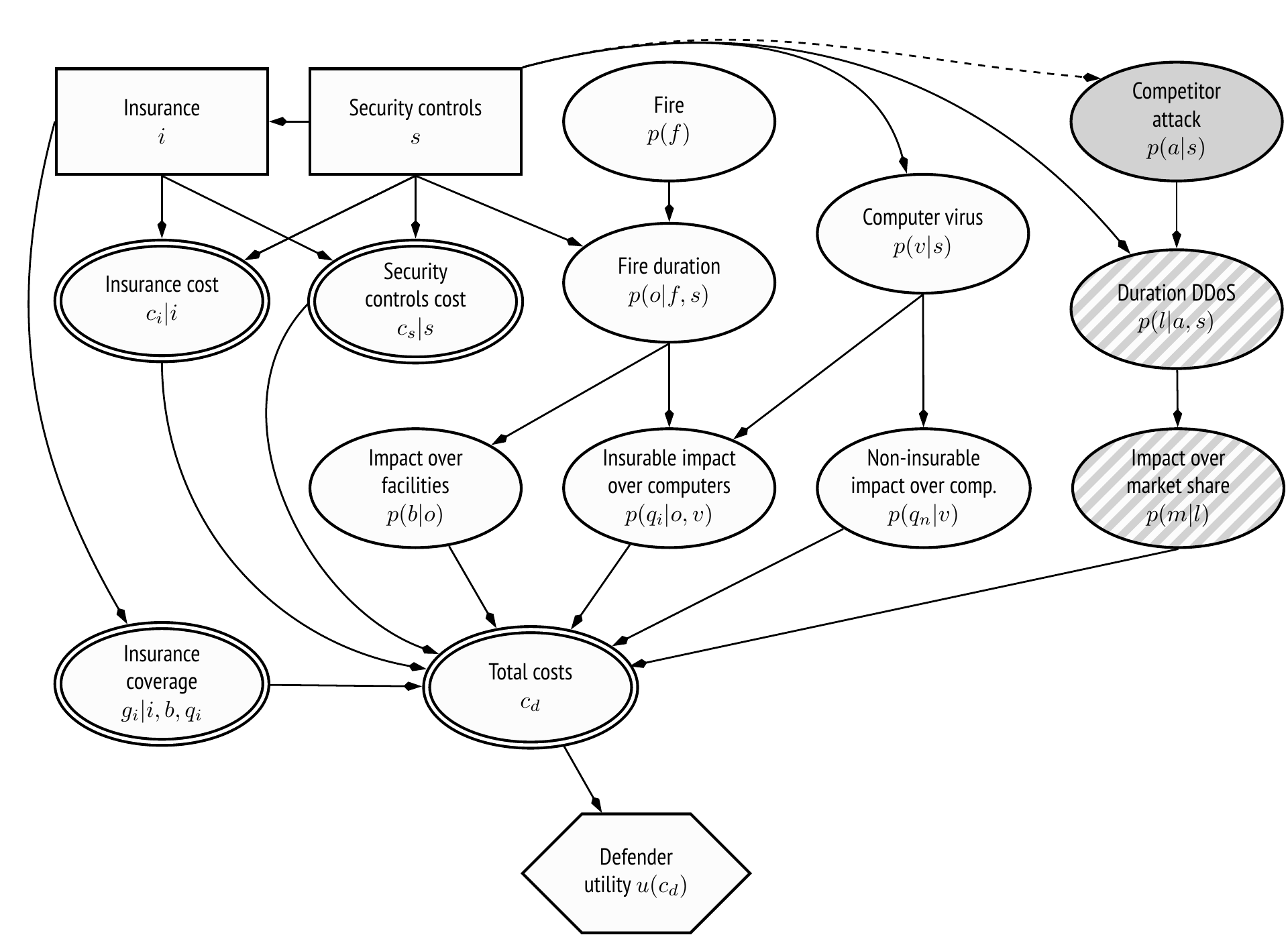}
	\caption{Defender problem.}
	\label{fig9a}
\end{figure*}

\begin{figure}
	\centering
	\includegraphics[scale=.7]{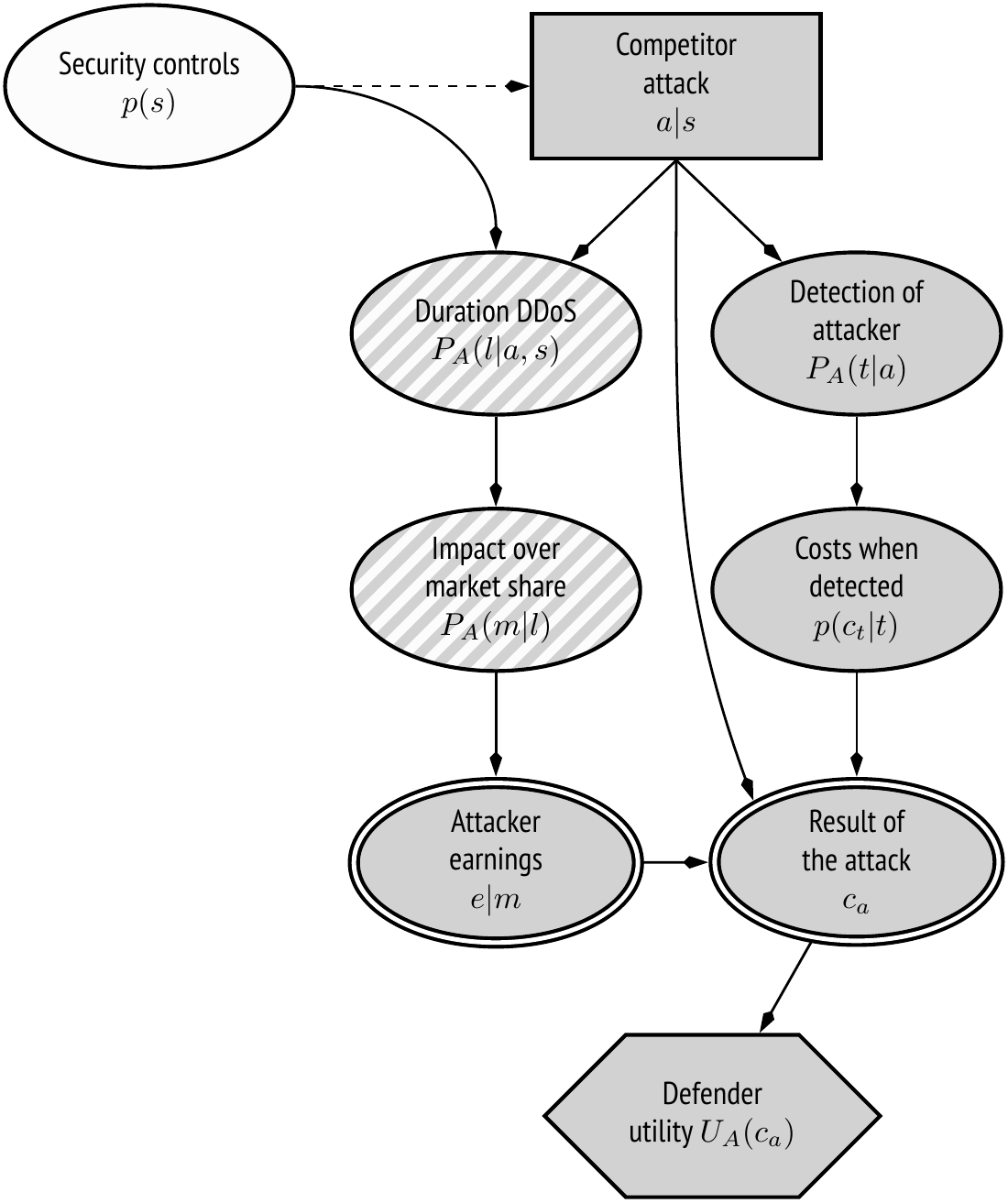}
	\caption{Attacker problem.}
	\label{fig9b}
\end{figure}

 Figs.~\ref{fig9a} and \ref{fig9b} respectively represent the Defender and Attacker problems derived from the strategic problem in Fig.~\ref{fig8}. For the Defender problem, this converts the Attacker's decision nodes into chance nodes and eliminates the Attacker's nodes not affecting the Defender problem, as well as the corresponding utility node. Similarly for the Attacker.
 We  use both diagrams to guide judgement elicitation from the Defender.

%%%%%%%%%%%%%%%%%%%%%%%%%%%%%%%%%%%%%%
\subsection{Assessing the Defender's non-strategic beliefs and preferences}
\label{subsec:defenderproblem}
We provide now the quantitative assessment of the Defender beliefs and preferences not requiring strategic analysis. Some of them will be based on data and expert judgement, others just on expert judgement due to the lack of data typical in many cybersecurity environments. As a consequence, we populate most nodes in her problem. We incorporate in Sect.~\ref{subsec:attackerproblem} those requiring strategic analysis. Finally, in Sect~\ref{subsec:simuldefender} we analyse the Defender problem to find the optimal controls and insurance. When incumbent, we provide the pertinent utility  $u()$, random utility $U_A()$, probability $p()$, random probability $P_A()$ or deterministic model at the corresponding node.

%%%%%%%%%%%%%%%%%%%%%%%%%%%%%%%%%%%%%%%%%%%%%%%%
\subsubsection{Economic value of Defender assets}	
We consider the following values for the assets at risk:
\begin{itemize}
	\item \emph{Facilities}: Their value is 5,000,000\euro, reflecting only acquisition costs.
	\item \emph{Computer equipment}: Valued at 200,000\euro, under similar considerations.
	\item \emph{Market share}: Currently estimated at 50\%. Translated into next year foreseen profits, we value it at 1,500,000\euro.
\end{itemize}

%%%%%%%%%%%%%%%%%%%%%%%%%%%%%%%%%%%%%%%%%%%%%%%
\subsubsection{Modelling security controls}
\label{subsubsec:modcontrols}

\paragraph{Security controls decision, $ s $:} The security portfolios that the Defender could implement derive from these options:
\begin{itemize}
	\item Install an anti-fire system.
	\item Install a firewall to protect the infrastructure.
	\item Train employees on safety and cybersecurity procedures.
	\item Subscribe a cloud-based DDoS protection system with choice (2, 5, 10 or 1000) gbps.
\end{itemize}

\noindent We thus have 40 portfolios. These could be further constrained by, e.g., a budget, as in Sect.~\ref{subsec:further}.

%%%%%%%%%%%%%%%%%%%%%%%%%%%%%%%%%%%%%%%%
\paragraph{Cost of security controls, $ c_s | s $:} This node models the cost of implemented controls. Table~\ref{table31} provides their costs, from which we derive the portfolio costs.

	\begin{table}
		\footnotesize
		\begin{center}
			\begin{tabular}{|l|c|}
				\hline
				\textbf{Security control} & \textbf{Cost} \\
				\hline
				Anti-fire system &   \textup{\euro} 1,500 \\
				\hline
				Firewall &   \textup{\euro} 2,250 \\
				\hline
				Risk mitigation proc. &   \textup{\euro} 2,000 \\
				\hline
				\parbox{10em}{Cloud-based DDoS \\ protection} & \parbox{11em}{\vspace{3pt} \textup{\euro} 2,400 for 2 gbps, \\ \textup{\euro} 3,600 for 5 gbps, \\ \textup{\euro} 4,800 for 10 gbps, \\ \textup{\euro} 12,000 for 1 tbps.} \\
				\hline
			\end{tabular}
			\caption{\label{table31} Cost of individual security controls.}
		\end{center}
	\end{table}

%%%%%%%%%%%%%%%%%%%%%%%%%%%%%%
\subsubsection{Modelling the insurance product}
\label{subsubsec:modinsurance}

\paragraph{Insurance decision, $ i $:} This refers to the insurance product that the Defender could purchase (Table~\ref{table32}) once the controls have been selected.

\paragraph{Insurance cost, $ c_i | i $:} This
 models the insurance premiums. It depends on the controls implemented by the organisation (Table~\ref{table32}).

\begin{table}
	\footnotesize
		\begin{center}
			\begin{tabular}{| l | c | c | c | c |}
				\hline
				\textbf{\multirow{2}{3em}{Prod.}} & \multicolumn{4}{ c | }{\textbf{Security controls}} \\
				\cline{2-5}
				& None & Anti-fire & \parbox{5em}{\vspace{3pt} \centering Firewall or \\ DDoS prot.} & Proc. \\
				\hline
				None & \textup{\euro} 0 & \textup{\euro} 0 & \textup{\euro} 0 & \textup{\euro} 0 \\
				\hline
				Trad. & \textup{\euro} 500 & \textup{\euro} 300 & \textup{\euro} 500 & \textup{\euro} 500 \\
				\hline
				Cyber & \textup{\euro} 300 & \textup{\euro} 300 & \textup{\euro} 200 & \textup{\euro} 250 \\
				\hline
				Compr. & \textup{\euro} 700 & \textup{\euro} 500 & \textup{\euro} 600 & \textup{\euro} 650 \\
				\hline
			\end{tabular}
			\caption{\label{table32} Insurance product cost.}
	\end{center}
\end{table}

%%%%%%%%%%%%%%%%%%%%%%%%%%%%
\paragraph{Insurance coverage, $ g_i | i, b, q_i $:} This node models $ g_i $, the insurance product coverage reflected in Table \ref{table2}. Traditional and comprehensive insurances cover 80\% of burnt facilities and computer costs. The cyber and comprehensive insurances will cover 80\% of the expenses related with virus removal.

%%%%%%%%%%%%%%%%%%%%%%%%%%%%%%%%%%%%
\subsubsection{Modelling fire risk}
\label{subsubsec:modfire}

\paragraph{Fire likelihood, $ p (f) $:} This node provides the annual probability of suffering a fire in our facility. We use data from the Vitoria fire brigade \cite{vitoria09}, concerning interventions on industrial buildings (Table~\ref{table3}).
\begin{center}
	\begin{table}
		\footnotesize
		\begin{center}
			\begin{tabular}{|l|c|c|}
				\hline
				\textbf{Year} & \textbf{Buildings} & \textbf{Fires} \\
				\hline
				2005 &   1220 &    32 \\
				\hline
				2006 &   1266 &   29 \\
				\hline
				2007 &   1320 &   30 \\
				\hline
				2008 &   1347 &   28 \\
				\hline
				2009 &   1314 &   28 \\
				\hline
			\end{tabular}
			\caption{\label{table3}   Industrial fire data in Vitoria (2005-2009).}
		\end{center}
	\end{table}
\end{center}
 The fire rate remains fairly stable over the years. We estimate the probability that an organization suffers a fire in a year using a beta-binomial model with prior $\beta e (1/2, 1/2)$. The posterior would be
\[
f | \textrm{data} \sim \beta e \Big(1/2 + \sum_{i=1}^{5} x_{i}, 1/2 + \sum_{i=1}^{5} (n_{i} - x_{i}) \Big) \equiv \beta e(147.5, 6320.5),
\]
where $x_{i}$ designates the number of fires affecting industrial buildings and $n_{i}$, that of such buildings in the $i-$th year, $i= 1,...,5$. Such distribution can be reasonably summarised through its posterior expectation, $\hat{p} = 0.022$, since the posterior variance is small. Its value is estimated from the probability that there are no fires, $ p(0) = 1 - \hat{p} = 0.978$. The number of fires can be approximated with a Poisson $\mathcal{P}(0.022)$ distribution.  However, we consider only the probability that one fire occurs, since probabilities beyond that are tiny ($p (f > 1) = 0.00024 $. Thus, the number $f$ of fires will follow
\[ f \sim \min [ 1, \mathcal{P}(0.022)]. \]

\begin{figure}
	\begin{center}
		\includegraphics[scale=0.5]{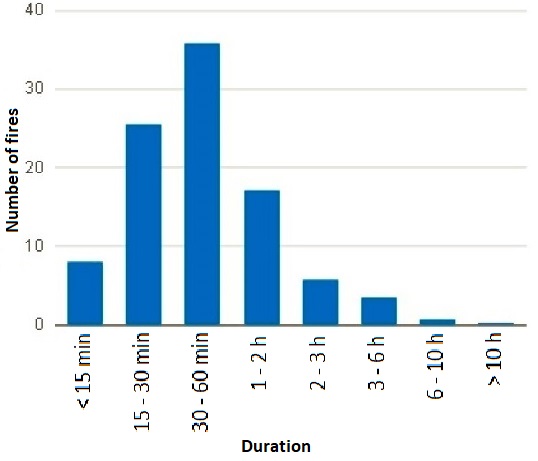}
		\caption{Industrial fire duration histogram. Vitoria, Spain (2005-2009).}
		\label{fig10} 
	\end{center}
\end{figure}

\paragraph{Fire duration, $ p ( o | f, s ) $:} It is a major impact determinant \cite{Bagchi2013}: the longer the fire, the more damaging it will be. To study its duration, we employ the above Vitoria data.
 Fig.~\ref{fig10} presents the histogram of industrial fire durations, with modal duration between 30 minutes and one hour.
Adopting the approach in \cite{wiper01}, we model fire duration with a gamma $ \Gamma (\textrm{shape}=\gamma, \textrm{scale}=\gamma/\mu) $
distribution. We assume a non-informative, but proper, prior $\gamma \sim Exp(0.01) $ and $\mu \sim \textrm{Inv-}\Gamma(1, 1) $. No analytical expression for the posterior distribution is available, but we can introduce a Markov chain MC scheme to sample from $\mu | data$ and $\gamma | data$, \cite{wiper01}. Based on this, we estimate that E($\gamma | data$) $\approx 0.85$ and E($\mu | data$) $\approx 78$.

The only security control among the proposed ones that may have an effect on fire duration is the anti-fire system, which enables faster fire detection. Using expert judgement \cite{dias2018}, we determine its threshold duration under the proposed system with, respectively, suggested minimum, modal and maximum durations of 1, 10 and 60 min. To mitigate expert overconfidence \cite{galway}, we consider a triangular distribution with quantiles 0.05 at 1 and 0.95 at 60 min, resulting in  $ Tri(0.8, 63, 10) $, which models the fire duration $ o $ if there is a fire ($f=1$) and portfolio $s$ contains the anti-fire system. On the other hand,
\[ o \sim \Gamma(0.85, 0.01089) \]
\noindent if the portfolio does not contain the anti-fire system.

%%%%%%%%%%%%%%%%%%%%%%%%%%%%%%%%%%%%%
\paragraph{Fire impact:} It models the impacts assuming that the fraction of affected assets is related with fire duration. After consulting with experts, we consider that a fire lasting 120 minutes would degrade the facilities by 100\% in absence of controls. To simplify, we assume that the effect of fire duration is linear. This impact will be assessed in Sect.~\ref{subsubsec:modimpact}.

Additionally, the impact over computer equipment derives from the percentage of facility degradation caused by fire. Assuming that computers are evenly distributed through the premises, a fire lasting 120 minutes would also degrade computer equipment by 100\%. This impact is potentially insurable and will be modelled in Sect.~\ref{subsubsec:modimpact}.

%%%%%%%%%%%%%%%%%%%%%%%%%%%%%%%%%%%%%%%%
\subsubsection{The computer virus risk}
\label{subsubsec:modvirus}

%%%%%%%%%%%%%%%%%%%%%%%%%%%%%%%%%%%%%%%
\paragraph{Computer virus likelihood, $ p ( v | s ) $:} This node provides the number of virus infections during a year in the organisation. The probability that a computer is infected in a month follows a binomial distribution $\mathcal{B}(m, q)$, where $q$ represents the probability that a computer gets infected. For infection duration, we assume that the virus remains active until detected through appropriate controls. Then, it becomes eradicated by the system administrator. Various statistics suggest that the rate of virus infections worldwide is 33\% \cite{Panda2015}, so we adopt $\hat{q} = 0.33 $ as the probability that a computer is infected a certain month. The organisation has 90 computers, which we assume have the same security controls and are equally likely of being infected. Since the analysis is for 12 months, we use $ m = 12 \cdot 90 = 1080 $.
Additionally, we consider the effect of our controls:
\begin{enumerate}
	\item If a firewall is implemented, the probability that a computer gets infected is reduced to $\hat{p} = 0.005$, not completely eliminating the threat, even if this includes continuous updating based on latest virus signatures.
	\item If maintenance is implemented, the infection probability gets reduced by 50\%, with firewall or not, as this control entails improvements in the organisation such as imposing safety requirements on acquired systems.
\end{enumerate}

\noindent The number $v$ of infections is, therefore, modelled as in Table~\ref{tablevirus}.
	\begin{table}[h]
		\renewcommand{\arraystretch}{1.5}
		\footnotesize
		\begin{center}
			\begin{tabular}{|l|c|}
				\hline
				\textbf{\parbox{9em}{Sec. controls in $s$}} & \textbf{Distribution} \\
				\hline
				Firewall and proc. & $ v \sim \mathcal{B} (1080, 0.0025) $ \\
				\hline
				Firewall & $ v \sim \mathcal{B} (1080, 0.005) $ \\
				\hline
				Procedure & $ v \sim \mathcal{B} (1080, 0.1666) $ \\
				\hline
				Otherwise & $ v \sim \mathcal{B} (1080, 0.33) $ \\
				\hline
			\end{tabular}
			\caption{\label{tablevirus} Number of annual virus infections.}
		\end{center}
		\renewcommand{\arraystretch}{1}
	\end{table}

%%%%%%%%%%%%%%%%%%%%%%%%%%%%%%%%
\paragraph{Computer virus impact:} Viruses may affect the three information security dimensions.The impact on integrity and availability could lead to information corrupting or unavailability. Impacts over confidentiality are variable, as they depend on the stolen information. The average daily cost of these infections was estimated at \textup{\euro} 2.683 \cite{Solutionary2013}, although this one varies according to the monetary value of the information and services the victim systems support. Bigger losses come from sophisticated campaigns (e.g., global ransomware like WannaCry) or targeted malware which, under our paradigm, should be better modelled as an adversarial threat. In our case, repairing a computer infected by a virus costs \textup{\euro} 31 (two technician hours). Insurance options potentially cover the removal of computer viruses. Therefore, this impact is modelled within the insurable aspects in Sect.~\ref{subsubsec:modimpact}.

Most computer viruses cause performance reduction in aspects such as initialisation of OS. Although small, this causes time losses to the user. We assume that most of the work time of the organisation is in front of a computer (70\%), and that it would take, on average, five days (40 h of work) to detect the problem. We also assume that when a computer is infected, 28 hours of its usage are affected by the virus. We model the time loss as a uniform distribution $ \mathcal{U}(0,0.05) $, representing that the percentage of lost time caused by a virus is between 0 and 5 \%. The hourly cost of the employee is \textup{\euro} 20/hour. Therefore, for each virus infection, the cost would be $ 20 \times 28 \times \mathcal{U}(0, 0.05) $. Insurance options in node $ i $ do not cover this loss and, thus, we modelled it within the non-insurable aspects in Sect.~\ref{subsubsec:modimpact}.

%%%%%%%%%%%%%%%%%%%%%%%%%%%%%%%%%%%%%%%
\subsubsection{Modelling the DDoS threat}
\label{subsubsec:modddosdef}
 We consider now non-strategic aspects of the DDoS threat. A model for the DDoS likelihood is in Sect.~\ref{subsec:attackerproblem}.

%%%%%%%%%%%%%%%%%%%%%%%%%%%%%%%%%%%%%%%%%%%%%%%%%%
\paragraph{DDoS Duration, $ p(l | a, s) $:} This node models the duration $l$ in hours of all successful DDoS attacks. Its length will depend on the intensity of the attacking campaign, how well crafted is the attack and the security controls implemented by the targeted organisation. Typical controls mitigating DDoS attacks involve configuring the digital system so that users and processes dedicate some resources for a certain period of time or distributing loads through a load balancer. An emerging alternative are cloud-based systems absorbing traffic from its customer site when they become victims of a DDoS. Otherwise, if no control is deployed, it would be virtually impossible to block such attack. Based on information in \cite{Securelist,verizonddos}, the average attack lasts 4 hours, averaging 1 gbps, with peaks of 10 gbps.
We model $ l_j $, the length of the $j$-th individual DDoS attack as a $ \Gamma(4, 1)$, so that its average duration is 4 hours. This duration is conditional on whether the attack actually saturates the target, which depends on the capability of the DDoS platform minus the absorption of the cloud-based system. We assume that the Attacker uses a professional platform capable of 5 gbps attacks, modelled through a $ \Gamma(5, 1)$ distribution. We then subtract the $ s_{\textrm{gbps}} $ absorbed by the protection system to determine whether the DDoS is successful, which happens when its traffic overflows the protection system. Since the campaign might take $ a $ attacks, the output of this node is
\[ l = \sum_j^a l_j, \]
\noindent with $ l_j \sim \Gamma(4,1) $ if $ \Gamma(5,1) - s_{\textrm{gbps}} > 0 $, and $ l_j = 0 $, otherwise.

%%%%%%%%%%%%%%%%%%%%%%%
\paragraph{DDoS impact:} The DDoS duration might cause a reputational loss that would affect the organisation market share. Recall that the current market share is 50\% valued at \textup{\euro} 1,500,000. To simplify, we assume that all market share is fully lost at a linear rate until lost in, say, 5-8 days of unavailability (120-192 hours of DDoS duration): in the fastest case the loss rate would be $ 0.5/120 = 0.00417 $ per hour, whereas in the slowest one it would be $0.0026 $. We model this with a uniform distribution $ \mathcal{U}(0.0026, 0.00417) $.

%%%%%%%%%%%%%%%%%%%%%%%%%%%%%%%%%%%%%%%%%%%%%%%%
\subsubsection{Modelling impacts over assets}
\label{subsubsec:modimpact}

\noindent We recall now the impacts over the assets.
%%%%%%%%%%%%%%%%%%%%%%%%%%%%%%%%
\paragraph{Impact over facilities, $ p(b|o) $:} This node models monetary losses $b$ due to the degradation of facilities by fire.
Following Sect.~\ref{subsubsec:modfire}, we model $b$ through
\[ b \sim 5000000 \times \min \Big( 1, \frac{o}{120} \Big) . \]

%%%%%%%%%%%%%%%%%%%%%%%%%%%%%%%%%%%%%%
\paragraph{Insurable impacts over computers, $ p(q_i | o, v) $:} This models the monetary losses $q_i$ due to degradation of computers to be covered by an insurance. This may be caused by fire, Sect.~\ref{subsubsec:modfire}, and by repairing the computers infected with viruses, Sect.~\ref{subsubsec:modvirus}. We then model $q_i$ through
\[ q_i \sim 31 v + 200000 \times \min \Big( 1, \frac{o}{120} \Big) . \]

%%%%%%%%%%%%%%%%%%%%%%%%%%%%%%%%%%%%%%%%%%%%%%%%%%%%%%
\paragraph{Non-insurable impacts over computers, $ p(q_n | v) $:} This models the monetary losses $q_m$ caused by degradation of computers
 not covered by insurance, due to the lost time caused by viruses over computer systems. Following Sect.~\ref{subsubsec:modvirus}, we model $q_n$ through
\[ q_n \sim 560 v \times \mathcal{U}(0,0.05)  .\]

%%%%%%%%%%%%%%%%%%%%%%%%%%%%%%%%%%
\paragraph{Impact over market share, $ p( m | l ) $:} This models the monetary value $m$ of market share lost. Following Sect.~\ref{subsubsec:modddosdef}, we use
\[ m \sim \min [1500000, l \times \mathcal{U} (0.0026, 0.00417)] . \]

%%%%%%%%%%%%%%%%%%%%%%%%%%%%%%%%%%%%%%%
\paragraph{Total costs for the Defender, $ c_d | g_i, c_i, c_s, m, b, q_i, q_n $:}
This models the costs $c_d$ suffered by the Defender through
\[ c_d = m + b + q_i + q_n + c_s + c_i - g_i , \]
being $c_s$ the cost of security controls, $c_i$ the cost of insurance, $g_i$ the insurance coverage (which reduces losses) and $m$, $b$, $q_i$ and $q_n$ the impacts over assets earlier described.

%%%%%%%%%%%%%%%%%%%%%%%%%%%%%%%%%%%%%
\subsubsection{Defender utility, $ u(c_d) $:}
The organisation is constant risk averse over costs. Its utility function is strategically equivalent to
\[ u(c_d) = a - b \exp(k(-c_d)), \]
We adjust it calibrating the function with three costs: worst, best, and an intermediate one. The worst reasonable loss $\max c_d $ is based on the sum of all costs and impacts (except the computer virus one) which is equal to \textup{\euro} 6,755,300. Computer virus impacts do not have an upper limit; based on simulations, it is reasonable to assume that they would not exceed \textup{\euro} 50,000. Giving an additional margin, we assume that $\max c_d = 7000000$. The best loss is $\min{c_d} = 0$. For an intermediate cost $c_d^*$, we find its probability equivalent  \cite{Ortega2017} $\alpha$ so that $u(c_d^*)= \alpha$. For instance, asking the company, we have $u(c_d^*= 2660000)\simeq  .5$. Additionally, we rescale  the costs to the (0,1) range through  $1 - \frac{c_d}{7000000}$. Then, the utility function is
\[ u(c_d) = \frac{1}{\textrm{e}-1} \Bigg[ \exp\Bigg(1 - \frac{c_d}{7000000}\Bigg) - 1 \Bigg] .  \]

%%%%%%%%%%%%%%%%%%%%%%%%%%%%%%%%%%%%%%%%%%%%55
\subsection{Assessing the Attacker's random beliefs and preferences}
\label{subsec:attackerproblem}
In the Defender problem, the competitor attack is a probabilistic node modelling the number of attacks launched by the Attacker, given the security controls implemented by the Defender. We model the Attacker problem and simulate it to forecast its actions to obtain the probability distribution.

We must estimate the probability that the Attacker executes the DDoS, given the  Defender controls implemented. For that, we consider his decision problem in Fig.~\ref{fig9b}. Its solution would give the Attacker's optimal action. However, as argued in Sect.~\ref{subsec:aracyber}, we model our uncertainty about his preferences and beliefs through random utilities and probabilities to find the random optimal attack.

%%%%%%%%%%%%%%%%%%%%%%%%%%%%%%%%%%%%%%%%%%%%%%
\subsubsection{Defender's security controls}
This node is probabilistic for the Attacker. However, we assume that he may observe through network exploration tools whether the Defender has implemented relevant controls against his attack.

%%%%%%%%%%%%%%%%%%%%%%%%%%%%%%%%%%%%%%%%%%%%%%%%%%
\subsubsection{Competitor attack decision: $ a | s $ }
 In the attacker problem, it is reflected in a decision node, modelling how many attacks (between 0 and 30) the DDoS campaign will consist of. Attackers usually give up once the attack has been mitigated and move onto the next target or try other disruption methods. However, when the sole objective is the victim, the Attacker might continue the campaign for several days, causing a pervasive impact. In our case, we assume that a DDoS platform would need a day to deploy their resources to launch a powerful and hidden DDoS.

%%%%%%%%%%%%%%%%%%%%%%%%%%%%%%%%%%%%%%%%%%%%%%%%%%%%
\subsubsection{Duration of the DDoS: $ P_A(l | a, s) $}
 We base our estimation on that of the Defender (Sect.~\ref{subsubsec:modddosdef}). The length of the $j$-th individual DDoS attack is modelled through a random gamma distribution $ \Gamma_{\textrm{length}} (\upsilon, \upsilon/\mu) $ with $ \upsilon \sim \mathcal{U}(3.6,4.8) $ and $ \upsilon/\mu \sim \mathcal{U}(0.8,1.2) $, so that we add uncertainty about the average duration (between 3 and 6 hours) and the dispersion. Similarly, the attack gbps are modelled through a random gamma distribution $ \Gamma_{\textrm{gbps}} (\omega, \omega/\eta) $ with $ \omega \sim \mathcal{U}(4.8,5.6) $ and $ \omega/\eta \sim U(0.8,1.2) $. Next, we subtract $ s_{\textrm{gbps}} $ to $ \Gamma_{\textrm{gbps}} $ to determine whether the DDoS is successful, which happens when its traffic overflows the protection system. As in Sect.~\ref{subsubsec:modddosdef}, the number $l$ of hours for which the site is unavailable during the campaign is modelled as
\[ l = \sum_j^a l_j, \]
with $ l_j \sim \Gamma_{\textrm{length}} $ if $ \Gamma_{\textrm{gbps}} - s_{\textrm{gbps}} > 0 $, and $ l_j = 0 $ otherwise.

%%%%%%%%%%%%%%%%%%%%%%%%%%%%%%%%%%%%%%%%%%%%%%%%%%%%%%%
\subsubsection{Impact over market share: $ P_A( m | l ) $}
 We base our estimation on that of the Defender (Sect.~\ref{subsubsec:modimpact}), adding some uncertainty around such assessment. The market share value and percentage are not affected by the uncertainty, as this information is available to both agents. However, we model uncertainty in the market loss rate so that the fastest one (5 days in the Defender problem) is between 4 and 6 days in the Attacker problem and the slowest one (8 for Defender) is between 7 and 9. Therefore, the random distribution describing the market loss $ m $ is
\[ m \sim \min \Big[ 1500000, l \times \mathcal{U}(\alpha, \beta) \Big] \]
\noindent with $ \alpha \sim \mathcal{U}(0.0021, 0.0031) $ and $ \beta \sim \mathcal{U}(0.00367, 0.00467) $.

%%%%%%%%%%%%%%%%%%%%%%%%
\subsubsection{Attacker earnings: $  e | m $}
This node models the Attacker gain $e$ in terms of market share, derived from the DDoS duration. As the sole competitor, we assume that $e$ corresponds to the share lost by the defender $ e = m $. The random uncertainty in the earnings is derived from the randomness of the preceding nodes.

%%%%%%%%%%%%%%%%%%%%%%%%%%%%
\subsubsection{Detection of Attacker: $ P_A(t | a) $}
 This node represents the chance of the Attacker being detected. In most cyber attacks, the attacker is not identified or prosecuted\footnote{For instance, the FBI Internet Crime Compliant Center prosecuted two cases, and investigated 73, of nearly  298,728 complaints received in 2016 \cite{2016fbi}}. Detection probabilities are estimated via expert judgement at 0.2\%, should the Attacker attempt a DDoS. He actually gambles his detection through a binomial distribution $ \mathcal{B}(a, 0.002) $ where the number of trials is the number $a$ of attacks and the detection probability is 0.002. To add some uncertainty, we model the detection probability for each attack through a beta distribution $ \beta e (2, 998) $\footnote{Its mean is 0.002}. Therefore, the distribution determining the detection of the attacker $ t $ is modelled through a random binomial distribution that produces the output \emph{detected} if $ \mathcal{B}(a, \phi) > 0 $ with $ \phi \sim \beta e (2, 998) $, and \emph{not detected}, otherwise.

%%%%%%%%%%%%%%%%%%%%%%%%%%%%%%
\subsubsection{Costs for Attacker when detected: $ p_A(c_t | t ) $}
 This node models the consequences associated with being detected when executing a DDoS. As a competitor, if the Attacker is disclosed, it would entail a serious discredit, together with compensation and legal costs besides incurring criminal responsibilities. To fix ideas, we use this cost decomposition:
\begin{itemize}
	\item Expected reputational costs, due to the necessary communication actions to preserve credibility: \textup{\euro} 550,000.
	\item Expected legal costs: \textup{\euro} 30,000.
	\item Expected civil indemnities and regulatory penalties: \textup{\euro} 350,000.
	\item Expected suspension costs, related with losses derived from prohibition to operate for some time: \textup{\euro} 1,500,000.
\end{itemize}

\noindent To add uncertainty, we model the costs as a normal distribution with mean 2430000 and standard deviation 400000, i.e.,
\[ c_t|t \sim \mathcal{N} (2430000, 400000) . \]

%%%%%%%%%%%%%%%%%%%%%%%%%%%%%
\subsubsection{Result of attack: $ c_a | e, c_t, a $ }

\noindent This node combines the attacker earnings and costs if detected, as well as the cost of undertaking the attacks. To estimate these, we consider that using a botnet to launch the DDoS attack would cost on average around \textup{\euro} 33 per hour \cite{Incapsula2015}. Each attack would take one day, entailing costs of \textup{\euro} 792. Therefore,
\[ c_a = e - c_t - 792 a .\]

%%%%%%%%%%%%%%%%%%%%%%%%%%%%%%%%
\subsubsection{Attacker's random utility: $ U_A(c_a) $}

\noindent We assume that the Attacker is risk prone, with utility function strategically equivalent to
\[u(c_{a}) = (c'_a)^{k_a} , \qquad k>1 ,\]

\noindent where $ c'_a $ are the costs $ c_a $ normalised to $ [0,1] $, and $ k_a $ the risk seeking attitude of the attacker. To induce uncertainty, we assume $k_{a}$ follows a $\mathcal{U}(8, 10)$ distribution.
Therefore, the attacker random utility is
\[ U_A(c_{a}) = (c'_a)^{K_a} \]
\noindent with $ K_a \sim \mathcal{U}(8,10) $.

%%%%%%%%%%%%%%%%%%%%%%%%%%%%%%%%%
\subsubsection{Simulating the Attacker problem}
\label{subsec:attackersolving}

\noindent Summarising the earlier assessments, the \emph{distribution of random utilities and probabilities in the Attacker problem} is
\[ F = \Big ( U_A ( c_a ), p_A ( c_t | t), P_A ( t | a ), P_A ( m | l), P_A ( l | a, s) \Big) . \]

\noindent We calculate the \emph{random optimal attack}, given the security controls $s$ implemented, as
\[
A^*(s) = \arg\max_{a} \int \dots \int \
U_A(c_a) \, p_A(c_t | t) \,  P_A(t | a) \, P_A(m | l) \, P_A(l | a, s) \
dl \, dm \, dt \, dc_t .
\]

\noindent To approximate it, we may use an MC approach as in Algorithm~\ref{algo1} (see Appendix), which we implemented in R. For each $ s $, the size of the DDoS protection system, we can assess the distribution of the random optimal attack. Table~\ref{attdistro} displays the probabilities of the attacks, conditional on the protection implemented, with $K=1000$. For instance, if the security portfolio contains no DDoS-protection system, an attack seems certain and its duration would be between 18 and 30 attacks (being 29 and 30 the most likely attack sizes). From this, we create the probability distribution $p(a|s)$, so that the Defender problem is fully specified and ready to be solved.

\begin{table*}[h]
	\renewcommand{\arraystretch}{1.2}
	\setlength\tabcolsep{2pt}
	\footnotesize
	\centering
	\begin{tabular}{|r|l|l|l|l|l|l|l|l|l|l|l|l|l|l|l|l|}
	\hline
	\multicolumn{1}{|c|}{\multirow{2}{*}{\textbf{\begin{tabular}[c]{@{}c@{}}DDoS\\ prot. system\end{tabular}}}} & \multicolumn{16}{c|}{\textbf{Number of attempts}}                                                                                                                                                                   \\ \cline{2-17} 
	\multicolumn{1}{|c|}{}                                                                                             & \textbf{0} & \textbf{1} & \textbf{2} & \textbf{3} & \textbf{4} & \textbf{5} & \textbf{6} & \textbf{7} & \textbf{8} & \textbf{9} & \textbf{10} & \textbf{11} & \textbf{12} & \textbf{13} & \textbf{14} & \textbf{15} \\ \hline
	\textbf{1 tbps}                                                                                                    & 1.000      & 0.000      & 0.000      & 0.000      & 0.000      & 0.000      & 0.000      & 0.000      & 0.000      & 0.000      & 0.000       & 0.000       & 0.000       & 0.000       & 0.000       & 0.000       \\ \hline
	\textbf{10 gbps}                                                                                                   & 0.000      & 0.001      & 0.003      & 0.003      & 0.004      & 0.005      & 0.012      & 0.012      & 0.015      & 0.013      & 0.017       & 0.024       & 0.024       & 0.022       & 0.030       & 0.035       \\ \hline
	\textbf{5 gbps}                                                                                                    & 0.000      & 0.000      & 0.000      & 0.000      & 0.000      & 0.000      & 0.000      & 0.000      & 0.000      & 0.000      & 0.000       & 0.000       & 0.001       & 0.001       & 0.001       & 0.002       \\ \hline
	\textbf{2gbps}                                                                                                     & 0.000      & 0.000      & 0.000      & 0.000      & 0.000      & 0.000      & 0.000      & 0.000      & 0.000      & 0.000      & 0.000       & 0.000       & 0.000       & 0.000       & 0.000       & 0.000       \\ \hline
	\textbf{none}                                                                                                      & 0.000      & 0.000      & 0.000      & 0.000      & 0.000      & 0.000      & 0.000      & 0.000      & 0.000      & 0.000      & 0.000       & 0.000       & 0.000       & 0.000       & 0.000       & 0.000       \\ \hline
	\end{tabular}
	\begin{tabular}{|r|l|l|l|l|l|l|l|l|l|l|l|l|l|l|l|}
	\hline
	\multicolumn{1}{|c|}{\multirow{2}{*}{\textbf{\begin{tabular}[c]{@{}c@{}}DDoS\\ prot. system\end{tabular}}}} & \multicolumn{15}{c|}{\textbf{Number of attempts}}                                                                                                                                                               \\ \cline{2-16} 
	\multicolumn{1}{|c|}{}                                                                                             & \textbf{16} & \textbf{17} & \textbf{18} & \textbf{19} & \textbf{20} & \textbf{21} & \textbf{22} & \textbf{23} & \textbf{24} & \textbf{25} & \textbf{26} & \textbf{27} & \textbf{28} & \textbf{29} & \textbf{30} \\ \hline
	\textbf{1 tbps}                                                                                                    & 0.000       & 0.000       & 0.000       & 0.000       & 0.000       & 0.000       & 0.000       & 0.000       & 0.000       & 0.000       & 0.000       & 0.000       & 0.000       & 0.000       & 0.000       \\ \hline
	\textbf{10 gbps}                                                                                                   & 0.026       & 0.041       & 0.025       & 0.044       & 0.042       & 0.053       & 0.050       & 0.048       & 0.047       & 0.060       & 0.050       & 0.059       & 0.065       & 0.081       & 0.089       \\ \hline
	\textbf{5 gbps}                                                                                                    & 0.008       & 0.006       & 0.012       & 0.017       & 0.007       & 0.028       & 0.031       & 0.055       & 0.070       & 0.061       & 0.096       & 0.117       & 0.143       & 0.141       & 0.203       \\ \hline
	\textbf{2gbps}                                                                                                     & 0.000       & 0.000       & 0.002       & 0.001       & 0.002       & 0.013       & 0.013       & 0.020       & 0.034       & 0.069       & 0.091       & 0.112       & 0.144       & 0.223       & 0.276       \\ \hline
	\textbf{none}                                                                                                      & 0.000       & 0.000       & 0.003       & 0.001       & 0.004       & 0.008       & 0.010       & 0.022       & 0.042       & 0.058       & 0.081       & 0.105       & 0.173       & 0.246       & 0.247       \\ \hline
	\end{tabular}
	\caption{\label{attdistro} Conditional probability table for random optimal attacks.}
\end{table*}

%%%%%%%%%%%%%%%%%%%%%%%%%%%%%%%%%%%%%%%%%%%%%
\subsection{Solution of the Defender problem}
\label{subsec:simuldefender}

\noindent Summarising earlier assessments about the Defender problem, the corresponding probabilities are
\[
G = \Big(p(m | l), p(q_n | v), p(q_i | o, v),
p(b | o), p(l | a, s), p(a | s), p(v | s), p(o | f,s), p(f) \Big) .
\]

\noindent The expected utility when the security portfolio $s$ is implemented together with insurance $i$ is
\begin{multline*}
\psi(s, i) = \int ... \int \
u( c_d ) \, p(m | l) \, p(q_n | v) \,  p(q_i | o, v) \, p(b | o) \, p(l | a, s ) \, p(a | s) \, p(v | s) \, p(o | f, s) \, p(f) \\
df \, do \, dv \, da \, dl \, db \, dq_i \, dq_n \, dm .
\end{multline*}

\noindent We can calculate the \emph{optimal allocation} as the maximum expected utility portfolio-insurance pair
\[(s^*, i^*) = \arg\max_{s , i} \psi (s, i).\]
We may use Algorithm~\ref{algo4} (see Appendix), to approximate the portfolio expected utilities and the optimal portfolio for the Defender. We have implemented it in R to calculate them (Table~\ref{table6c}). Specifically, \emph{the best portfolio} consists of:
\begin{itemize}
	\item 1 tpbs cloud-based DDoS protection system.
	\item Firewall.
	\item Anti-fire system.
	\item Comprehensive insurance.
\end{itemize}

\begin{table*}[h]
	{\footnotesize
		\begin{center}
			\begin{tabular}{| l | l | l | l | l | c |}
				\hline
				\textbf{Anti-fire}
				& \textbf{Firewall}
				& \textbf{Procedure}
				& \textbf{DDoS protection}
				& \textbf{Insurance}
				& \textbf{Expected utility} \\
				\hline
				anti-fire & firewall & no procedure & 1 tbps & comprehensive & 0.9954 \\
				\hline
				anti-fire & firewall & no procedure & 1 tbps & traditional & 0.9950 \\
				\hline
				no anti-fire & firewall & no procedure & 1 tbps & comprehensive & 0.9949 \\
				\hline
				\dots & \dots & \dots & \dots & \dots & \dots \\
				\hline
				no anti-fire & no firewall & no procedure & no protection & no insurance & 0.8246 \\
				\hline
				no anti-fire & firewall & no procedure & no protection & cyber & 0.8246 \\
				\hline
				anti-fire & no firewall & no procedure & no protection & no insurance & 0.8242 \\
				\hline
			\end{tabular}
			\caption{\label{table6c} Expected utility for 3 best and worst combinations of controls and insurance.}
	\end{center}}
\end{table*}

\noindent Besides the ranking of countermeasures, we can obtain additional information from the simulation. For instance, the best security controls contain a firewall, 1 tbps DDoS protection and no risk management procedure. Additionally, the best portfolios also includes insurance, either traditional or comprehensive.

%%%%%%%%%%%%%%%%%%%%%%%%%%%%%%%%%%
\subsection{Further analysis}
\label{subsec:further}

\noindent The previous ARA model can be used to perform other relevant analysis, as we briefly discuss.
%%%%%%%%%%%%%%%%%%%%%%%%%%%%%%%%%%%%
\subsubsection{Sensitivity analysis}

\noindent By introducing variations in the probabilities (e.g., probability of fire), we can evaluate the robustness of the previous solution by checking whether variations in the probabilities and parameters of the model alter the optimal solution or the relevance of different controls. This is specially relevant in a case like ours with little differences in expected utility among top alternatives and many inputs are purely of judgemental nature. The approach would require the implementation of additional algorithms for sensitivity analysis that indicate whether a small deviation in a parameter may lead to a large effect in the outcome of the model \cite{Rios1990}. Additionally, sensitivity analysis can be used to explore the maximum cyber insurance price that the Defender would be willing to pay. This may be used to price insurance products, as well as for finding the best portfolio for different cybersecurity budgets.

%%%%%%%%%%%%%%%%%%%%%%%%%%%%%%%%%%%%%%%%%%%%
\subsubsection{Introducing constraints}

\noindent As we mentioned, we may introduce constraints over the security portfolios. For example, we could add to the problem a budget limit of, say, \textup{\euro}15,000. Then, our problem could involve only those portfolios satisfying such constraint. We can also consider constraints of insurance on security controls as in insurance policies there are some requirements regarding controls that the company should comply with to be insured. Other types of constraints could be dealt with similarly.

%%%%%%%%%%%%%%%%%%%%%%%%%%%%%%%%%%%%%%%%%%%%%
\subsubsection{Return on security investment}

\noindent Our formulation focused on choosing the best portfolio, but an additional aspect that could be addressed with our model is calculating the return on security investment (ROSI) to assess the cost effectiveness of a cybersecurity budget \cite{enisaROSI,Schatz2017}. Calculating the optimal solution over a range of budgets (e.g., from \textup{\euro}5,0000 to \textup{\euro}25,000) allows generating a function that, for a given budget, gives the optimal solution and expected utility to explore the return on risk mitigation investments. Additionally, we could find the optimal increase in the portfolio so as to attain a certain expected utility level or reach a certain risk appetite level.

%%%%%%%%%%%%%%%%%%%%%%%%%%%%%%%%%%%%%%%%%%%%%
\section{Discussion}
\label{sec:discussion}

\noindent Current cybersecurity risk analysis frameworks provide a thorough knowledge base for understanding cyber threats, security policies and impacts over assets that depend on the digital infrastructure. However, such frameworks provide risk analysis methods that are not sufficiently formalised, neither comprehensive enough. Most of them suggest risk matrices as their main analytic basis, which provide a fast but frequently rudimentary study of risks.

Hence, we present an ARA framework providing a formal method supporting all steps relevant to undertake a comprehensive cybersecurity risk analysis. It implies structuring the cybersecurity problem as a decision model based on a multiagent influence diagrams. ARA enables the assessment of beliefs and preferences of the organisation regarding cybersecurity risks as well as the security portfolio and insurance they can implement to treat such risks. It takes into account, in addition to non-intentional threats, the strategic behaviour of adversarial threats. We model the intentional factor through the decision problems of the Attackers. The case introduced is a simplification of a real example but serves as a template for other cases. Among other things, we had to rely on expert judgement for the uncertainty nodes for which we lacked data.

From the decision-making point of view, ARA enables the calculation of optimal cybersecurity resource allocations, facilitating the selection of security and insurance portfolios. Furthermore, it also enables sensitivity analysis to evaluate whether the optimal portfolio remains as optimal, in case different elements affecting risk change. This may be used for insurance pricing.

Future work involves the application of this paradigm to study other cybersecurity adversarial problems. The proposed problem refers to strategic/tactical decisions; it would be interesting to develop
dynamic schemes integrating strategic and operational decisions. Similarly, we shall address the development of parametric cyber insurance schemes, aimed at supporting the obtention of premiums that, as complement of the implemented controls, facilitate more effective risk management. Another relevant activity would be the development of a software environment that supports the implementation of the ARA framework for cybersecurity based on the R routines developed, as well as optimisation algorithms beyond enumeration, to reduce computational burden.

When compared with standard approaches in cybersecurity, our paradigm entails a more comprehensive method leading to a more detailed modelling of risk problems, yet more demanding in terms of analysis. We believe, though, that at many organisations, especially, critical infrastructures and sectors, the stakes at play are so high that the entailed additional work should be worth the effort.

%%%%%%%%%%%%%%%%%%%%%%%%%%%%%%%%%%%%%%%%%%%%%
%\section*{Acknowledgements}
%\noindent Work supported by the EU's Horizon 2020 project 740920 CYBECO (Supporting Cyberinsurance from a Behavioural Choice Perspective). The work of DRI is supported by the Spanish Ministry of Economy and Innovation program MTM2014-56949-C3-1-R, the ESF-COST Action IS1304 on Expert Judgement and the AXA-ICMAT Chair on Adversarial Risk Analysis.

\renewcommand{\refname}{REFERENCES}

\newpage

%%%%%%%%%%%%%%%%%%%%%%%%%%%%%%%%%%%%%%%%%%%%%
\section*{Appendix}
\begin{algorithm}[!htp]
	\caption{\label{algo1} Estimating distribution over attacks (defence-attack).}
	\ForEach{defence $e$}{
		\For{$i=1, \ldots, K$}{
			Generate %\vspace*{-1em}
			\[
			\Big( U_A^i (t_2, c_t, c_c), P_A^i (c_t | t_1, t_2, e), \\ P_A^i (c_c | t_1, t_2, e), P_A^i (t_1 | e) \Big) \sim F
			\]
			Compute %\vspace*{-1em}
			\[
			t_2^i = \arg\max_{t_2} \iiint U_A^i (t_2, c_t, c_c) \, P_A^i (c_t | t_1, t_2, e) \, P_A^i (c_c | t_1, t_2, e) \, P_A^i (t_1 | e) \\ dt_1 \, dc_c \, dc_t
			\]
		}
		Approximate %\vspace*{-1em}
		\[\hat{p}_A (t_2 | e) = \frac{\# \{ t_2^i = t_2 \} }{K}\]
	}
\end{algorithm}

\begin{algorithm}[!htp]
		\caption{\label{algo4} Approximation of Defender's optimal portfolio.}
		$ \psi(s, i) = 0 $ \\
		\ForEach{$(s,i)$} {
			\For{$j=1, \ldots, 1000$}{
				\noindent Generate %\vspace*{-1em} 
				\[ \big( m^j, q_n^j, q_i^j, b^j, l^j, a^j, v^j, o^j, f^j \big) \sim G \]
				Compute %\vspace*{-1em} 
				\[ c_s^j | s , \qquad
				c_i^j | i , \qquad
				g_i^j |  i, b^j, q_i^j \]
				Compute %\vspace*{-1em} 
				\[ c_d^j = m^j + b^j + q_i^j + q_n^j + c_s^j + c_i^j - g_i^j \]
				Compute %\vspace*{-1em} 
				\[ \psi(s, i) = \psi(s, i) + \frac{u(c_d^j)}{1000} \]
			}
		}
		Approximate %\vspace*{-1em} 
		\[ (\hat{s}^*, \hat{i}^*) = \arg\max_{s, i} \psi (s, i)  \]
		\vspace*{-1em} 
\end{algorithm}

\end{document}